%% file: bare_jrnl_new_sample4.tex
\documentclass[lettersize,journal]{IEEEtran}
\usepackage{amsmath,amssymb,amsfonts}
\usepackage{algorithmic}
\usepackage{algorithm}
\usepackage{array}
\usepackage[caption=false,font=normalsize,labelfont=sf,textfont=sf]{subfig}
\usepackage{cite}
\usepackage{graphicx}
\usepackage{xcolor}
\usepackage{colortbl}
\definecolor{catBlue}{RGB}{232, 241, 250}
\definecolor{catPurple}{RGB}{246, 238, 249}
\definecolor{catGreen}{RGB}{235, 247, 238}
\definecolor{catOrange}{RGB}{255, 244, 230}
\definecolor{tablecategory}{RGB}{240, 243, 248}
\usepackage{hyperref}
\usepackage{textcomp}
\usepackage{url}
\usepackage{verbatim}
\usepackage{multirow}
\usepackage{booktabs}
\usepackage{threeparttable}
\usepackage{float}
\begin{document}

\title{DeltaGateNet: Bidirectional Temporal Dynamics Modeling \\for
EEG-based Driving Fatigue Recognition}

\author{Yip Tin Po*, Jianming Wang*, Yutao Miao, Jiayan Zhang, Yunxu Zhao, \\Xiaomin Ouyang\textsuperscript{$\dagger$}, Zhihong Li\textsuperscript{$\dagger$}, Nevin L. Zhang\textsuperscript{$\dagger$}%
\thanks{*Equal contribution.}%
\thanks{$\dagger$ Xiaomin Ouyang, Zhihong Li, Nevin L. Zhang are the corresponding authors.}}

\markboth{IEEE Transactions on Intelligent Transportation Systems}
{Po \MakeLowercase{\textit{et al.}}: DeltaGateNet: Bidirectional Temporal Dynamics Modeling for EEG-based Driving Fatigue Recognition}


\maketitle
\begin{abstract}
    Driving fatigue is a leading cause of traffic accidents, posing critical challenges to the safety and operational reliability of intelligent transportation systems (ITS). While electroencephalography (EEG) provides direct neural correlates of vigilance, its practical application in vehicular environments is hindered by asymmetric neural dynamics and pronounced inter-subject variability. To address these challenges, we propose DeltaGateNet, a lightweight and computationally efficient framework for real-time EEG-based fatigue recognition in intelligent vehicular systems. Our architecture introduces a Bidirectional Delta module that explicitly decomposes first-order temporal differences into positive and negative components, capturing asymmetric neural activation and suppression patterns associated with vigilance degradation. Subsequently, a Gated Temporal Convolution module employs temporal convolutions and residual learning to robustly model channel-wise temporal dependencies while preserving spatial specificity. Extensive experiments on the three-class SEED-VIG and two-class SADT datasets under both intra-subject and inter-subject paradigms demonstrate that DeltaGateNet consistently outperforms state-of-the-art baselines. Moreover, the model directly ingests segmented time-domain EEG epochs without requiring multi-stage runtime filtering or artifact removal pipelines. These results demonstrate DeltaGateNet's promise for edge integration in driver monitoring systems, offering a computationally efficient, generalizable, and physiologically grounded solution for vigilance recognition in ITS.
\end{abstract}

\section{Introduction}
\IEEEPARstart{D}{riving} fatigue is widely recognized as one of the leading contributors to traffic accidents worldwide, particularly in long-duration and monotonous driving scenarios \cite{shaik2023systematic}. Recent large-scale studies have shown that fatigue-related impairment can substantially increase the risk of severe traffic accidents, posing a significant threat to public safety and transportation systems \cite{sprajcer2023tired}. Unlike alcohol or distraction, fatigue is a latent and gradually accumulated cognitive state, which makes it difficult to detect reliably through self-report or post-event analysis. Therefore, developing an objective, real-time, and reliable driving fatigue recognition system remains a critical yet unresolved challenge. Within the emerging paradigm of Transportation Cyber-Physical Systems (T-CPS), continuous physiological monitoring serves as a critical bridge between human operators and intelligent vehicular infrastructures, enabling proactive safety interventions before cognitive degradation compromises vehicle control. By eliminating the need for complex runtime handcrafted filtering, DeltaGateNet facilitates integration into in-vehicle monitoring frameworks for driver assistance and safety intervention.

Existing driving fatigue detection approaches can be broadly categorized into non-physiological and physiological methods \cite{al2024technologies}. Non-physiological techniques, such as behavioral analysis, vehicle dynamics monitoring, and vision-based methods, are often sensitive to environmental conditions and provide limited interpretability of the driver’s internal cognitive state \cite{ajayi2025multimodal,tian2024illumination}. In contrast, physiological approaches—particularly those based on electroencephalography (EEG)—offer a more direct and objective measurement of neural activity associated with vigilance and fatigue. However, many EEG-based methods still rely on handcrafted spectral features or static representations, which inadequately characterize the temporal evolution of fatigue-related neural processes \cite{yu2025application,cao2026set,zhou2025connectivity}. Although recent deep learning models have demonstrated improved representation learning capability, most existing architectures primarily focus on amplitude-based features or generic temporal modeling, while largely overlooking the bidirectional nature of neural dynamics and the inherent channel-wise independence of EEG recordings \cite{hu2025adaptive,wang2025eegmamba}.

The previously mentioned constraints arise mainly from the complex temporal properties of EEG data, which present considerable obstacles to reliable driving fatigue identification. Firstly, EEG data exhibit significant non-stationarity, with fatigue presenting as slow and irregular temporal variations instead of sudden state transitions \cite{gao2025idsrn}. Secondly, fatigue-associated brain mechanisms demonstrate significant bidirectional asymmetry, since reductions in arousal and compensatory neural activation do not adhere to symmetrical temporal patterns \cite{krigolson2025using,babu2019compensatory}. Thirdly, practical driving applications often depend on a restricted number of EEG channels, which limits spatial information and requires a stronger focus on robust channel-specific temporal modeling \cite{liu2024two,shalash2021deep}. Together, these factors limit the efficiency and generalizability of current EEG-based fatigue identification techniques, especially in cross-subject application scenarios.

Drawing from previous studies on non-stationary EEG modeling, asymmetric brain processing, and temporal convolutional networks, the key insight is that driving fatigue is more accurately defined by the temporal evolution of neural activity than by its absolute amplitude \cite{gao2025idsrn}. Bidirectional temporal variations in EEG signals offer essential insights into the decline of focus and brain inhibition mechanisms. Based on these insights, we propose a new framework called DeltaGateNet that directly models bidirectional temporal dynamics using a Bidirectional Delta module and captures channel-wise temporal dependencies using a Gated Temporal Convolution module. The Bidirectional Delta module decomposes first-order time differences into positive and negative components, allowing the network to differentiate between asymmetric neuronal activation and suppression patterns. The Gated Temporal Convolution module tracks temporal dynamics for each EEG channel through temporal convolutions and residual learning, maintaining spatial distinctiveness while extracting representative temporal features.

Comprehensive studies performed on three public driving fatigue datasets, SEED-VIG and SADT, in both intra-subject and inter-subject evaluation settings demonstrate the effectiveness and robustness of the proposed DeltaGateNet. DeltaGateNet achieves an intra-subject accuracy of 81.89\% and an inter-subject accuracy of 55.55\% on the three-class SEED-VIG dataset (a 6.64\% improvement over the strongest baseline). In the balanced SADT 2022 dataset, the proposed method obtains intra-subject and inter-subject accuracies of 96.81\% and 83.21\%, respectively. Similar performance is found on the unbalanced SADT 2952 dataset, with DeltaGateNet reaching 96.84\% intra-subject accuracy and 84.49\% inter-subject accuracy. While cross-subject generalization remains inherently challenging for multi-class fatigue prediction, these results demonstrate that explicitly modeling bidirectional temporal dynamics and channel-wise temporal dependencies facilitates robust and generalizable detection of driving fatigue across diverse subject and class distributions.

The main contributions of this work are summarized as follows:
\begin{itemize}
\item{We introduce DeltaGateNet, a novel framework for EEG-based driving fatigue recognition designed to explicitly model bidirectional temporal dynamics. By leveraging a Bidirectional Delta module to capture neural fluctuations and a Gated Temporal Convolution module for channel-wise dependencies, the architecture achieves robust representations when operating directly on segmented time-domain EEG epochs without requiring multi-stage runtime filtering pipelines. This end-to-end design provides a computationally efficient solution for driver vigilance monitoring.}

\item{We conducted extensive experiments on three benchmark driving fatigue datasets, SEED-VIG, SADT 2022, and SADT 2952, employing both intra-subject and strict subject-disjoint inter-subject evaluation paradigms. We demonstrate that DeltaGateNet consistently outperforms state-of-the-art baselines across both balanced and unbalanced class distributions, with detailed ablation and parameter sensitivity analyses.}

\item{To facilitate reproducibility and future research, the DeltaGateNet implementation is made publicly available as an open-source project at \url{https://github.com/yip-tin-po-alex/DeltaGateNet}. The code has been published.}
\end{itemize}

\section{Related Works}
Driving fatigue recognition based on EEG is inherently a temporal modeling problem, as fatigue reflects a gradual, accumulative, and often asymmetric degradation of cognitive states over time. From a modeling perspective, existing end-to-end EEG-based driving fatigue recognition methods can be systematically categorized according to how temporal dynamics are represented and constrained. From this perspective, previous research can be categorized into four paradigms: (1) implicit temporal aggregation, (2) multi-scale symmetric temporal modeling, (3) global temporal dependency modeling, and (4) lightweight and physiologically limited temporal modeling.

\subsection{Implicit Temporal Aggregation}
Early deep learning methodologies for EEG-based driving fatigue detection mostly utilize convolutional neural networks that implicitly capture temporal information by using fixed-size temporal kernels and pooling processes. In these methods, temporal dynamics are not explicitly modeled but are instead embedded as latent features during network optimization. Representative studies employ covariance-based CNNs to capture inter-channel relationships from EEG epochs \cite{parekkattil2024eeg}, or compact convolutional architectures to extract fatigue-related patterns from single-channel or low-density EEG recordings \cite{cui2022compact}. These designs reduce reliance on handcrafted elements and exhibit notable performance, especially in intra-subject evaluation contexts.

However, implicit temporal aggregation assumes that fatigue-related EEG features may be sufficiently captured within constrained temporal receptive fields. This assumption is limiting for driving fatigue, which develops continually and nonlinearly throughout extended driving durations. Thus, models that depend exclusively on implicit temporal encoding frequently demonstrate constrained sensitivity to subtle vigilance transitions and reduced robustness in cross-subject evaluations.

\subsection{Multi-Scale Symmetric Temporal Modeling}
To mitigate the constraints of fixed temporal receptive fields, further research introduces explicit multi-scale temporal modeling methodologies. Fully convolutional networks and Inception-style architectures utilize parallel temporal convolutions with diverse kernel sizes to capture EEG fluctuations at multiple temporal resolutions \cite{sun2025novel,tang2024attention}. InceptionTime and its associated versions illustrate this paradigm by concurrently simulating short-term fluctuations and long-term temporal patterns. Residual temporal convolutional networks improve training stability and feature propagation through skip connections, exhibiting robust performance in various time-series classification tasks \cite{sun2025novel,wang2025research}.

While multi-scale temporal modeling improves the representation of heterogeneous temporal patterns, these methods predominantly treat temporal variations in a symmetric manner, focusing on amplitude-based feature extraction. As a result, they fail to explicitly capture the bidirectional characteristics of temporal evolution, such as progressive enhancement or attenuation of neural activity, which are closely associated with fatigue accumulation and vigilance decline.

\subsection{Global Temporal Dependency Modeling}
More recently, attention-based architectures, including Transformer and Vision Transformer variants, have been introduced to EEG analysis to explicitly model long-range temporal dependencies. By leveraging self-attention mechanisms, these models can capture interactions across distant time points and channels. Hybrid CNN–Transformer frameworks combine local convolutional feature extraction with global attention modeling \cite{yang2025lgformer}, while ViT-based approaches reinterpret EEG signals as sequences of patches or time–frequency representations \cite{modesitt2024fusing}.

Despite their expressive power, global dependency models generally require large-scale datasets and incur substantial computational overhead. Moreover, most existing attention-based approaches aim to learn generic temporal relationships without explicitly encoding the bidirectionality of temporal changes or preserving channel-wise temporal independence. These limitations reduce their practicality for EEG-based driving fatigue recognition, particularly in real-world scenarios involving limited-channel wearable devices and constrained data availability.

\subsection{Lightweight and Physiologically Constrained Temporal Modeling}
In parallel, a line of research focuses on lightweight and physiologically motivated EEG networks that balance performance, interpretability, and computational efficiency. EEGNet is a representative example that decouples spatial and temporal feature learning through depthwise and separable convolutions, achieving competitive performance with significantly reduced model complexity \cite{lawhern2018eegnet}. Subsequent studies adapt such compact architectures to fatigue-related tasks by adjusting temporal kernel sizes, normalization strategies, and pooling operations \cite{hana2025enhanced}.

Although these models are well suited for real-time and wearable applications, their temporal modeling capacity remains largely implicit and constrained by predefined architectural assumptions. In particular, they do not explicitly model asymmetric or bidirectional temporal dynamics inherent in prolonged driving fatigue, limiting their ability to robustly characterize non-stationary vigilance degradation.

\section{Methods}
\begin{figure*}[t]  
    \centering
    \includegraphics[width=\linewidth]{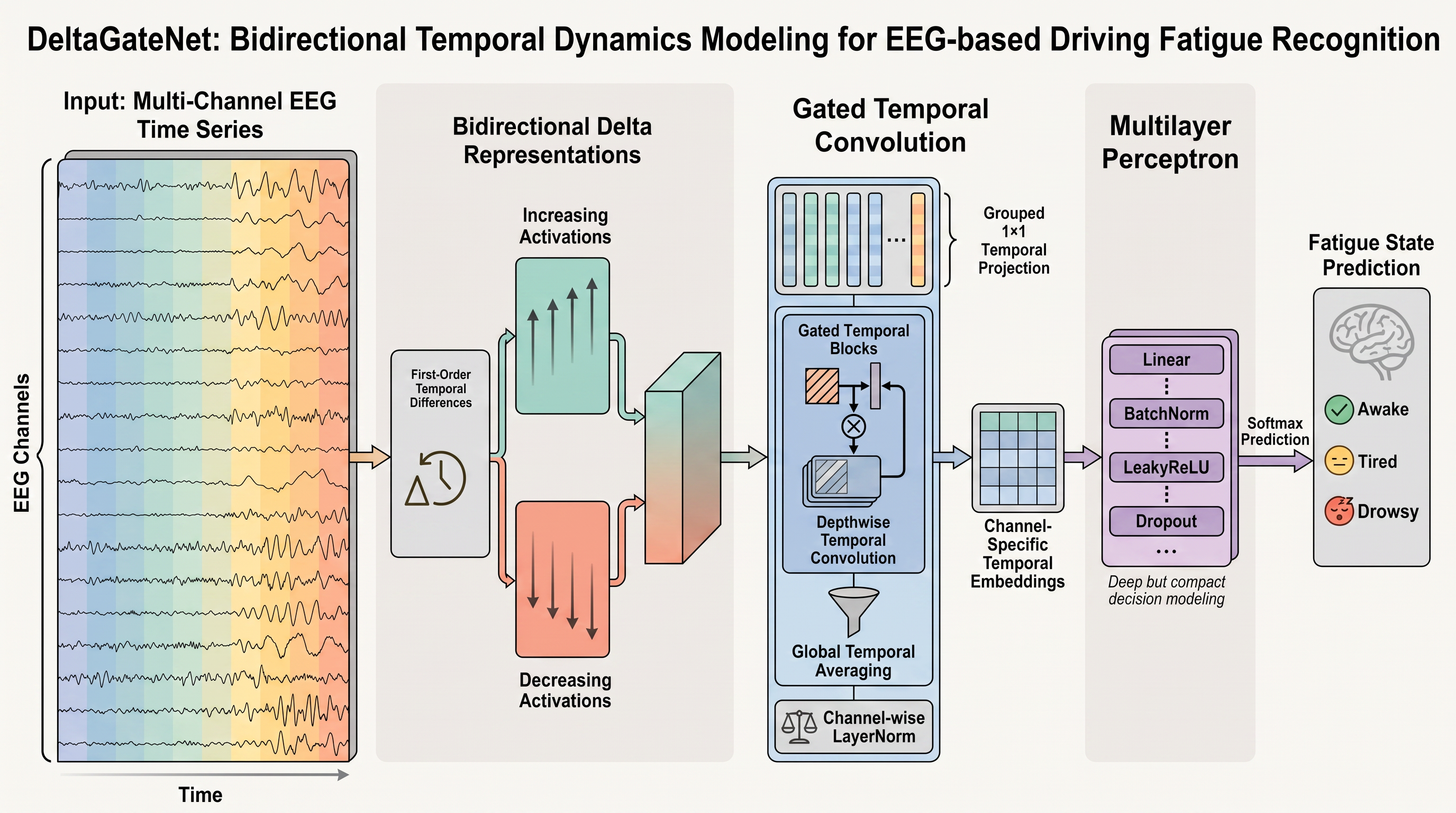} 
    \caption{Overview of the proposed DeltaGateNet. DeltaGateNet consists of three main components: (1) The Bidirectional Delta module computes first-order temporal differences across timesteps and separates positive and negative fluctuations via half-wave rectification. (2) The Gated Temporal Convolution module extracts channel-wise temporal features using temporal convolutions, GELU activation gating, and residual refinement. (3) The Multilayer Perceptron (MLP) classification head maps pooled temporal embeddings to output class probabilities.}
    \label{fig:overview} 
\end{figure*}

In this section, we present the architecture of DeltaGateNet for driving fatigue recognition on segmented time-domain EEG signals. The overall architecture of DeltaGateNet is illustrated in Fig. \ref{fig:overview}. The framework comprises three principal components: (1) a parameter-free Bidirectional Delta module that computes directional first-order temporal differences to capture rapid neural state transitions; (2) a Gated Temporal Convolution module that models channel-wise temporal patterns via temporal convolutions, GELU activation gating, and residual learning; and (3) a Multilayer Perceptron (MLP) classification head that maps the resulting temporal embeddings to class probabilities.

\subsection{Bidirectional Delta Representations}
Given an input EEG epoch $X \in \mathbb{R}^{B \times C \times T}$, where $B$ denotes the batch size, $C$ is the number of EEG electrode channels, and $T$ is the number of temporal sampling points, this module extracts the directional first-order discrete difference for each channel. For each channel sequence $x(t)$, the temporal difference $\Delta x(t)$ is defined with a causal boundary condition at $t=0$:
\begin{equation}
\Delta x(t) = 
\begin{cases} 
0, & t = 0 \\ 
x(t) - x(t-1), & t \in \{1, 2, \dots, T-1\} 
\end{cases}
\label{eq:delta}
\end{equation}
The zero-padding at the initial timestep preserves the original sequence length $T$ while strictly preventing any future temporal information from leaking into preceding time points. 

To explicitly capture asymmetric neural dynamics associated with cognitive state transitions, the temporal difference is decomposed into positive and negative components using half-wave rectification:
\begin{equation}
\Delta x^+(t) = \text{ReLU}(\Delta x(t)) = \max(0, \Delta x(t))
\label{eq:pos_delta}
\end{equation}
\begin{equation}
\Delta x^-(t) = \text{ReLU}(-\Delta x(t)) = \max(0, -\Delta x(t))
\label{eq:neg_delta}
\end{equation}
where $\Delta x^+(t)$ isolates rising phases of neural fluctuations (activation), and $\Delta x^-(t)$ isolates falling phases (suppression). 

The positive and negative difference tensors $\Delta X^+, \Delta X^- \in \mathbb{R}^{B \times C \times T}$ are concatenated along the channel dimension to yield a bidirectional differential representation:
\begin{equation}
X_{\text{bidir}} = [\Delta X^+; \Delta X^-] \in \mathbb{R}^{B \times 2C \times T}
\label{eq:bidir_concat}
\end{equation}
By providing the network with decoupled rising and falling components, the model directly accesses the directionality and rate of change in neural oscillations without increasing learnable parameters \cite{Sound Event Differential,Rectified Derivative}. This allows downstream layers to learn distinct filters for oscillatory rise and decay phases, producing representations that are more robust to inter-subject amplitude variations.

\subsection{Gated Temporal Convolution}
Following the Bidirectional Delta module, the tensor $X_{\text{bidir}} \in \mathbb{R}^{B \times 2C \times T}$ is processed by the Gated Temporal Convolution module to capture multi-scale temporal dependencies for each individual EEG channel.

First, a $1 \times 1$ convolution projects the $2C$ input channels into a higher-dimensional feature space:
\begin{equation}
X_0 = \text{Conv1D}_{1\times 1}(X_{\text{bidir}}) \in \mathbb{R}^{B \times 2CD \times T}
\label{eq:proj}
\end{equation}
where $D$ denotes the hidden feature depth per channel. This expands the representational capacity of each electrode channel independently.

Next, the model employs stacked residual blocks to learn temporal dynamics while stabilizing gradient flow. Each residual block consists of a 1D temporal convolution with $\text{groups}=1$ and kernel size $K = 7$. 

The convolution is followed by Batch Normalization (BatchNorm) and a Gaussian Error Linear Unit (GELU) activation function \cite{GELU}, which provides smooth nonlinear gating. 

The gated activations are then processed by a $1 \times 1$ pointwise convolution to mix features across depth channels, followed by Dropout for regularization. A residual connection is added around the block:
\begin{equation}
X_l = X_{l-1} + \mathcal{F}(X_{l-1})
\label{eq:residual}
\end{equation}
where $X_{l-1}$ and $X_l$ denote the input and output tensors of the $l$-th residual block, and $\mathcal{F}(\cdot)$ represents the composite operation of temporal convolution ($\text{groups}=1$), BatchNorm, GELU gating, pointwise convolution, and Dropout. The residual connection facilitates the propagation of both high-frequency transient features and deeper abstracted representations \cite{resid_1,resid_2}.

After passing through two residual blocks ($l=1, 2$), the tensor $X_2 \in \mathbb{R}^{B \times 2CD \times T}$ is reshaped to $\mathbb{R}^{B \times 2C \times D \times T}$. Temporal average pooling is applied along the time dimension to yield a compact $D$-dimensional temporal embedding vector $\bar{X} \in \mathbb{R}^{B \times 2C \times D}$ for each of the $2C$ bidirectional channels. Finally, channel-wise Layer Normalization is applied across the $D$ dimensions to standardize the learned embeddings.

\subsection{Multilayer Perceptron Classification Head}
The pooled and normalized feature tensor $\bar{X} \in \mathbb{R}^{B \times 2C \times D}$ is flattened into a one-dimensional feature vector of dimension $2CD$. The vector is fed into a 3-layer Multilayer Perceptron (MLP). Each of the first two layers comprises a Linear transformation, Batch Normalization, LeakyReLU activation ($\alpha = 0.01$), and Dropout ($p = 0.5$). The final Linear layer projects the hidden representations to $k$ output logits corresponding to the $k$ target classes ($k=3$ for SEED-VIG and $k=2$ for SADT). The Softmax operation is applied to convert logits into predicted class probabilities $\hat{y} \in \mathbb{R}^{B \times k}$, and the entire network is trained end-to-end using standard cross-entropy loss.

\section{Experimental Design}
\subsection{Datasets and Input Pipelines}
Three publicly available EEG benchmarks for driver vigilance estimation are evaluated in this study: the SEED-VIG dataset \cite{Seed vig} and the Sustained-Attention Driving Task (SADT) dataset \cite{sadt paper}.

\subsubsection{SEED-VIG}
The SEED-VIG dataset \cite{Seed vig} contains multi-channel EEG recordings collected in a simulated virtual driving environment from 23 subjects (11 male, 12 female; mean age 23.3 years). Each subject underwent a continuous driving trial lasting approximately 120 minutes. EEG signals were acquired from 17 electrode channels positioned according to the international 10--20 system at a sampling rate of 200~Hz. 

Vigilance levels were continuously tracked using PERCLOS (percentage of eyelid closure over the pupil) \cite{Perclos} captured via eye-tracking glasses. Following standard protocol \cite{Seed vig}, continuous recordings were segmented into non-overlapping 8-second epochs ($T = 1600$ time points), resulting in 885 samples per subject trial. Ground-truth vigilance labels were categorized into three discrete classes based on PERCLOS ($p$): Alert ($p \in (0, 0.35]$), Semi-Fatigue ($p \in (0.35, 0.7]$), and Fatigue ($p \in (0.7, 1.0]$). The public SEED-VIG dataset was bandpass filtered ($0.5$--$50$~Hz) by the dataset curators, and our framework ingests these 17-channel time-domain epochs directly without requiring additional runtime feature extraction or handcrafted artifact screening.

\subsubsection{Sustained-Attention Driving Task (SADT)}
The SADT dataset \cite{sadt paper} evaluates driver drowsiness during sustained lane-keeping tasks in a driving simulator. Drivers maintained the vehicle within the center lane while unpredictable lane-deviation events occurred periodically, prompting drivers to execute quick corrective steering maneuvers. Driver state was labeled based on the local reaction time (RT) between deviation onset and the corrective maneuver (deviations with RT significantly exceeding baseline alert RT are designated as drowsy, while normal response times are designated as alert).

\begin{figure}[htbp]
    \centering
    \includegraphics[width=\columnwidth]{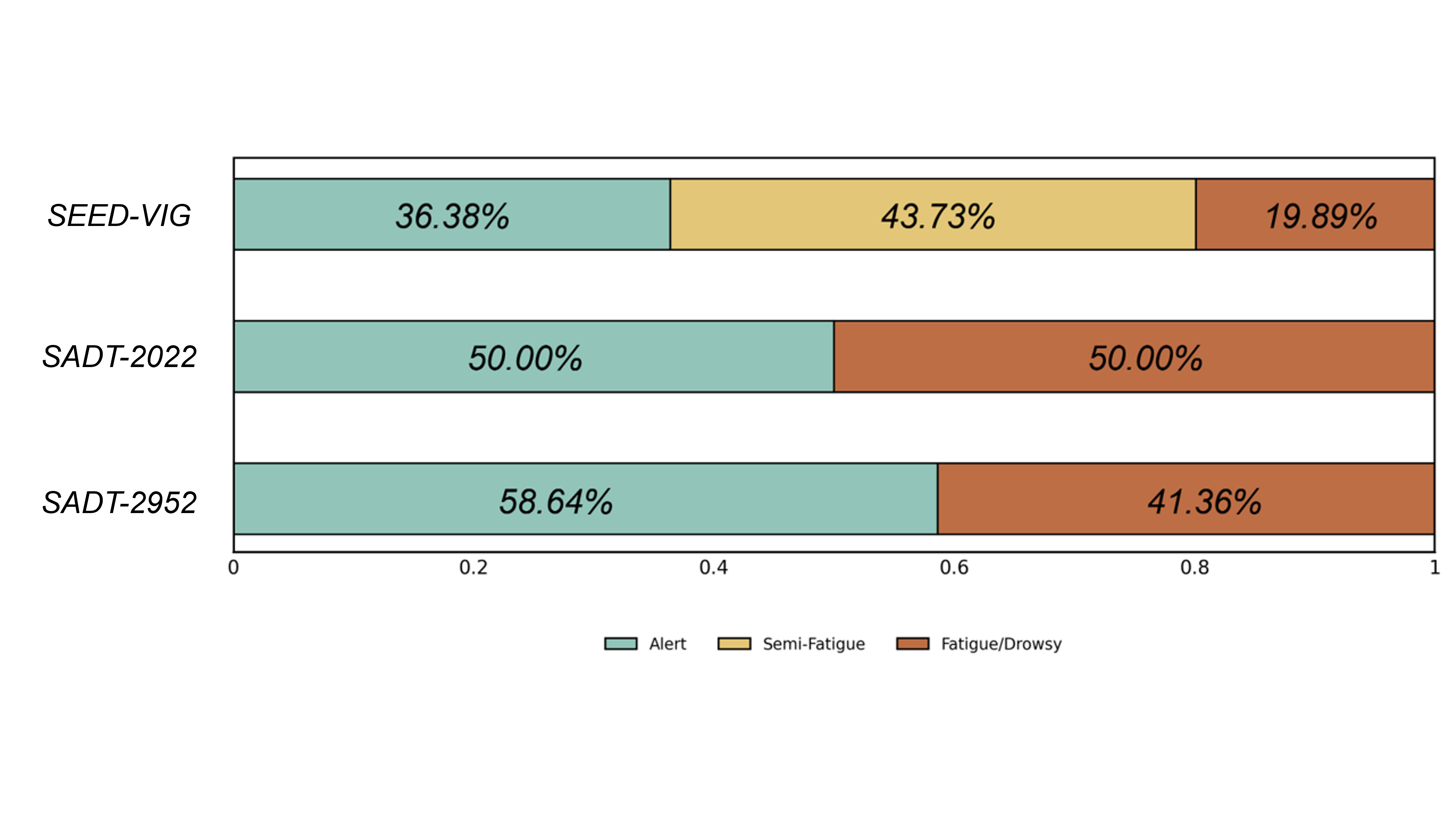}
    \caption{Class distribution of the three benchmark datasets. SEED-VIG is a three-class dataset (Alert, Semi-Fatigue, Fatigue), while SADT 2022 and SADT 2952 are binary (Alert, Drowsy).}
    \label{fig:class_distribution}
\end{figure}

We evaluate both benchmark releases of SADT derived from 11 subjects with 30-channel EEG recorded at 128~Hz. The SADT 2022 release \cite{sadt 2022} is a balanced subset containing 2,022 three-second epochs ($T = 384$ time points) preceding deviation events, with an equal distribution between alert and drowsy states. The SADT 2952 release \cite{sadt 2952} is the full unbalanced dataset containing 2,952 three-second epochs cross the 11 subjects, reflecting natural class imbalance. Fig.~\ref{fig:class_distribution} summarizes the class distribution across the three datasets. Detailed per-class counts and percentages are tabulated in Appendix~\ref{sec:appendix_class_distribution}.

\subsection{Data Splitting and Leakage Prevention}
\begin{figure}[htbp]
    \centering 
    \includegraphics[width=0.5\textwidth]{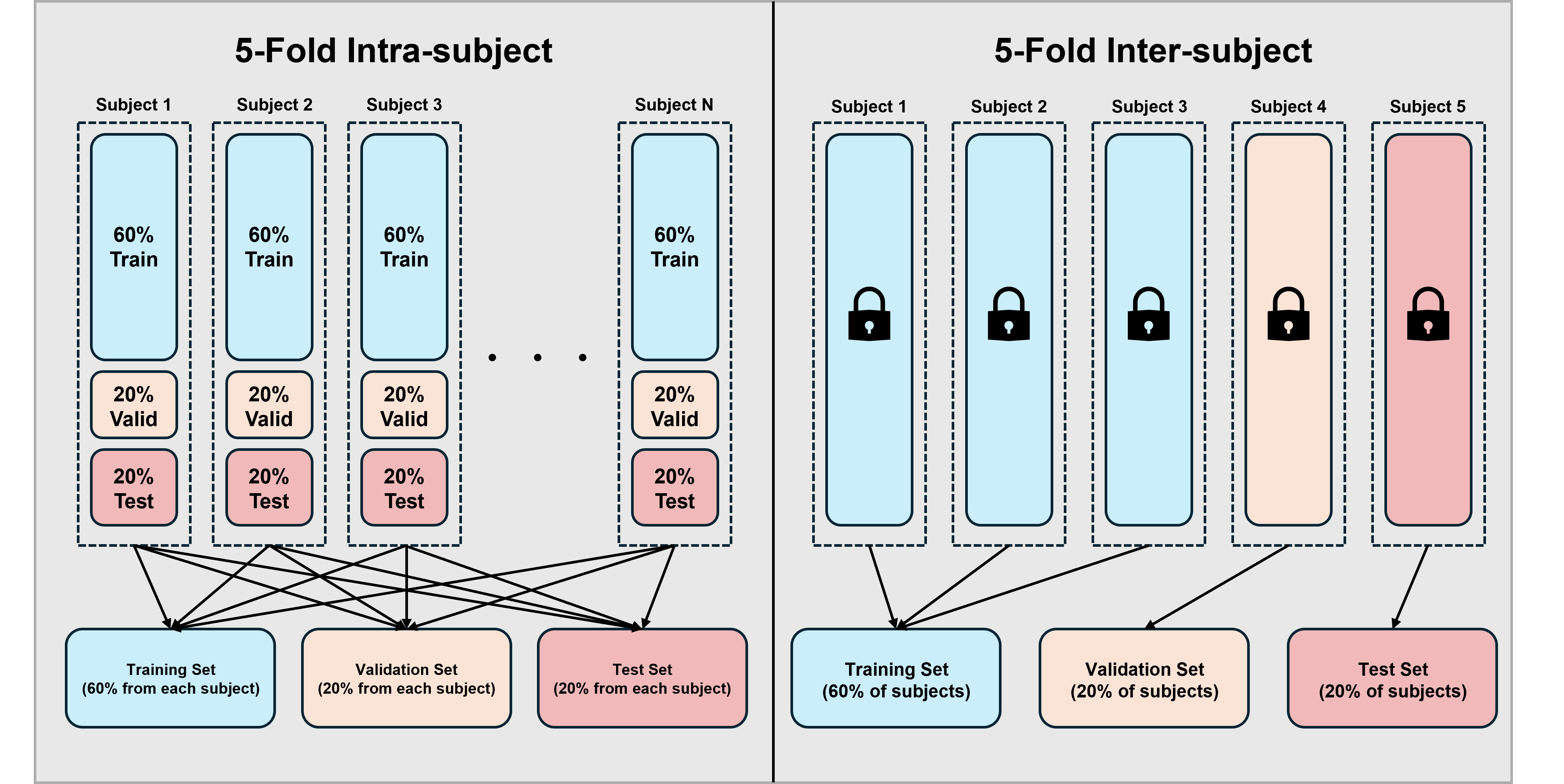} 
     \caption{Intra-subject and inter-subject 5-fold evaluation splits. Intra-subject folds partition each subject's epochs into 60\%/20\%/20\% train/validation/test sets; inter-subject folds assign mutually exclusive subject groups to the three partitions.}
    \label{fig:Split Visualization} 
\end{figure}
To rigorously evaluate model performance and prevent data leakage, we establish evaluation protocols for both intra-subject and inter-subject paradigms using 5-fold cross-validation. 

In the intra-subject evaluation setting, for each subject, temporal EEG epochs are partitioned into 5 folds using a 60\% training, 20\% validation, and 20\% testing split. This tests the model's capacity to recognize fatigue states for a calibrated, known individual.

In the inter-subject (cross-subject) evaluation setting, to assess generalization to previously unseen individuals, we enforce a strict subject-disjoint partitioning scheme. For the 23 subjects in SEED-VIG and the 11 subjects in SADT, subjects are grouped into mutually exclusive subsets across the 5 folds (approximately 60\% training subjects, 20\% validation subjects, and 20\% test subjects per fold). No subject appears in more than one partition within any fold, ensuring zero cross-subject data leakage. All models are trained exclusively on training subjects, tuned on validation subjects, and evaluated on unseen test subjects. Figure \ref{fig:Split Visualization} is a visual representation of our evaluation settings.

\subsection{Baselines}
For a comprehensive analysis of our model's performance, we compared DeltaGateNet against ten benchmark architectures spanning time series classification models, CNN-Transformer-based models, EEG-specific models, and EEG foundation models.

\subsubsection{Time Series Classification Models}
FCN\_Wang \cite{Resnet_Wang} is a benchmark architecture constructed from three sequential convolution blocks, utilizing kernel sizes of 8, 5, and 3, respectively. InceptionTime \cite{InceptionTime} utilizes six stacked inception modules that integrate parallel convolutional pathways with four distinct kernel sizes. Resnet\_Wang \cite{Resnet_Wang} is a 1D residual network architecture incorporating three 3-layer residual blocks with larger kernel sizes of 64, 128, and 256.

\subsubsection{Deep Learning CNN and Transformer-Based Models}
XResnet \cite{XResnet101} is a modified ResNet architecture incorporating a 2$\times$2 average pooling operation with a stride of 2 in the downsampling pathway to preserve information during dimensionality reduction. Vision Transformer (ViT) \cite{ViT} models sequential patches using the Transformer multi-head self-attention mechanism to capture global visual and temporal relationships.

\subsubsection{EEG-Specific Models}
EEGNet \cite{EEGNet} employs depthwise and separable convolutions to construct a compact spatial-temporal network architecture tailored for diverse EEG classification tasks. Modified TSception \cite{TSception} enhances spatiotemporal feature extraction through a multi-scale temporal refinement strategy and adaptive average pooling for cognitive state assessment. GAT-CNN \cite{GATCNN} integrates Graph Attention Networks with CNNs, utilizing a mutual information-driven connectivity matrix to model inter-channel dependencies for driving fatigue detection.

\subsubsection{EEG Foundation Models}
Large EEG foundation models aim to extract generalized, transferable features across diverse electrode layouts and recording paradigms. We benchmark against two fine-tuned foundation models: LaBraM \cite{LaBraM}, which tokenizes continuous EEG signals into discrete neural codes via vector-quantized neural spectrum prediction, and EEGMamba \cite{EEGMamba}, which introduces a state space model backbone with spatial-temporal adaptation modules for EEG sequence modeling.

\subsection{Implementation Details and Evaluation Metrics}
All models were implemented in PyTorch 2.9.0 and executed on NVIDIA A100 GPUs. Optimization was performed using the AdamW optimizer with an initial learning rate of $1 \times 10^{-4}$ ($1 \times 10^{-5}$ for LaBraM fine-tuning), weight decay of $1 \times 10^{-2}$, batch size of 32, and 200 maximum training epochs. A fixed random seed ($\text{seed} = 2026$) was initialized across all runs to guarantee deterministic splits and complete reproducibility. Every baseline ingests the same segmented time-domain EEG tensors as DeltaGateNet (SEED-VIG: $17 \times 1600$; SADT: $30 \times 384$) with no additional filtering or artifact-cleaning stage, and is trained and evaluated on the identical 5-fold splits under this shared optimization budget. Model performance is quantified using four standard evaluation metrics: Accuracy (Acc), Macro-Averaged Precision (Pre), Macro-Averaged Recall (Rec), and Macro-Averaged F1-Score (F1). Reporting macro-averaged metrics ensures fair evaluation under class imbalance. Macro-F1 equally weights per-class precision and recall and therefore plays the same imbalance-aware role as balanced accuracy (macro-averaged recall).

\section{Experimental Results}
\label{sec:experimental_results}
\begin{figure}[htbp]
    \centering
    \includegraphics[width=\columnwidth]{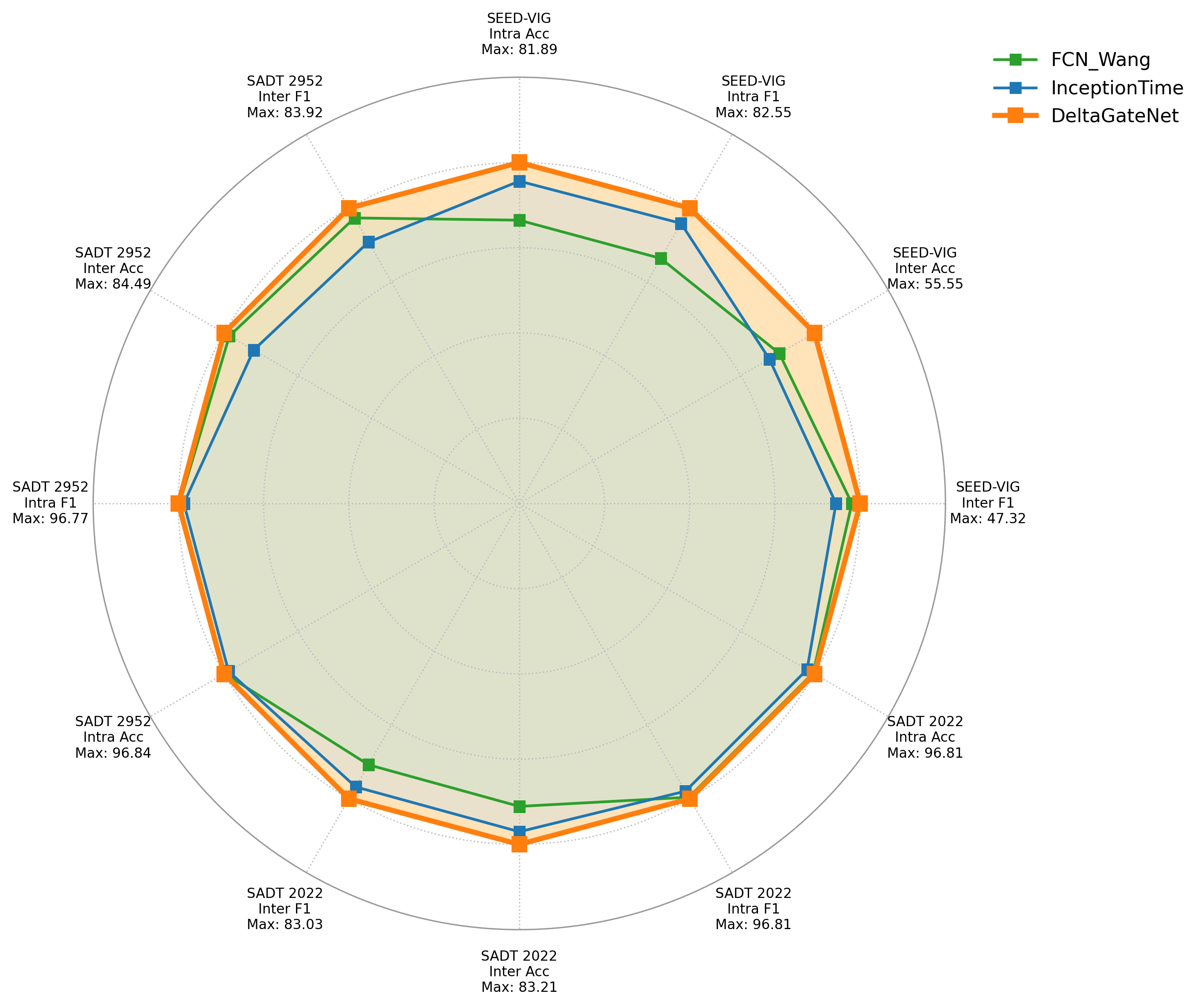}
    \caption{Radar chart comparing DeltaGateNet with the strongest baselines across all datasets and evaluation settings. Each of the 12 axes corresponds to one (dataset, evaluation setting, metric) combination, spanning SEED-VIG, SADT 2022, and SADT 2952 under both intra-subject and inter-subject paradigms for Accuracy and Macro-F1. Each axis is normalized by the best score across models.}
    \label{fig:radar}
\end{figure}

This section details the evaluation of our proposed model, DeltaGateNet. Performance was assessed through five-fold intra-subject and inter-subject experiments, comparing it against baseline methods from four categories: time series classification models, deep learning CNN/Transformer-based models, established EEG-specific models, and EEG Foundation Models. Ablation studies were also conducted to quantify the contribution of each model component and to measure the significance of the Bidirectional Delta module. All baseline models were trained and evaluated on the exact same cross-validation splits and input tensors using identical optimization criteria to guarantee strict evaluation fairness. Fig.~\ref{fig:radar} compares DeltaGateNet with the strongest baselines across all datasets, settings, and metrics.

\subsection{Comparison Among Previous Models}
\subsubsection{Intra-Subject Experiments}
Table~\ref{tab:intra_subject_results} presents the intra-subject classification performance, evaluated through 5-fold cross-validation, across the SEED-VIG and two SADT datasets for all compared models. The proposed DeltaGateNet consistently achieves the highest performance across all four evaluation metrics on every dataset.

On the SEED-VIG dataset, DeltaGateNet attains an Accuracy of \(81.89 \pm 0.66\%\), surpassing the next best method (TSception, \(79.80 \pm 0.98\%\)) by nearly 3 percentage points. It also shows substantial gains in Precision (\(84.37\%\) vs. \(81.26\%\)), Recall (\(81.32\%\) vs. \(80.29\%\)), and F1-score (\(82.55\%\) vs. \(80.68\%\)). Moreover, DeltaGateNet achieves an 8\% increase over EEGNet on accuracy.

On the SADT 2022 dataset and the SADT 2952 dataset, DeltaGateNet surpasses all SOTA models. Notably, time series classification models outperform both EEG-based models and transformer-based models on multiple datasets. 

Among baseline models, time series classification methods generally perform well, with InceptionTime and Resnet\_Wang being strong competitors. However, DeltaGateNet maintains a clear and consistent advantage across nearly all metrics, demonstrating its effectiveness in intra-subject fatigue recognition. In comparison, transformer and deep-learning convolution-based models underperform time series models. This suggests that EEG data to measure driving fatigue is not directly correlated with the size or depth of the model. 

\input{Intrasubject}
\subsubsection{Inter-Subject Experiments}
Table~\ref{tab:inter_subject_results} presents the more challenging inter-subject classification results, evaluated through 5-fold cross-subject validation, across the SEED-VIG and SADT datasets. In this rigorous cross-subject paradigm, our proposed DeltaGateNet demonstrates stronger generalization by consistently achieving the highest accuracy on all three datasets.

On SEED-VIG, DeltaGateNet achieves \(55.55 \pm 3.96\%\) accuracy, outperforming the strongest baseline (FCN\_Wang, \(48.91 \pm 8.05\%\)) by approximately 6.64 percentage points. While cross-subject variance remains high across all evaluated models due to individual physiological differences in multi-class fatigue manifestation, DeltaGateNet exhibits superior resilience.

On SADT 2022, DeltaGateNet attains \(83.21 \pm 6.25\%\) accuracy, surpassing InceptionTime (\(80.12 \pm 9.11\%\)) by over 3 percentage points with lower variance. ViT ranks next among the baselines; DeltaGateNet surpasses it by nearly 8 percentage points in accuracy, suggesting stronger generalization than both time-series classification models and transformer-based models. 
On SADT 2952, DeltaGateNet achieves \(84.49 \pm 7.36\%\) Accuracy, exceeding FCN\_Wang (\(82.99 \pm 11.61\%\)) while showing greater stability with smaller variance under class imbalance. 

As observed, the margin between first and second place is consistently wider on the SEED-VIG dataset than on SADT. This stems from a fundamental difference in task difficulty: SADT is a binary classification problem, while SEED-VIG is a three-class classification problem. This challenging paradigm serves to better demonstrate the representation learning capabilities of DeltaGateNet.

These results confirm that DeltaGateNet not only performs well in intra-subject evaluation but also generalizes robustly to unseen subjects, outperforming both specialized EEG models and general time-series classifiers.

\input{Intersubject}

\subsection{Ablation Experiments}
\input{Ablation}
\input{Ablation_Delta}
\subsubsection{Influence of Each Individual Module}
In this section, we investigate the individual contributions of the core architectural components of DeltaGateNet. Ablation experiments were conducted across all three benchmark datasets under both intra-subject and inter-subject evaluation paradigms, systematically removing modules to assess their impact. The results are summarized in Table~\ref{tab:combined_ablation}. As observed across all evaluations, removing either the Bidirectional Delta module or the Gated Temporal Convolution module leads to substantial performance degradation, confirming the essential role of both components.

On SEED-VIG (Table~\ref{tab:combined_ablation}), in the intra-subject setting, the base model without both modules (MLP only) achieves an accuracy of $43.72\%$. Removing the Bidirectional Delta module (w/o Bidirectional Delta) drops accuracy from $81.89\%$ to $71.01\%$ (a $10.88\%$ reduction), demonstrating that the parameter-free directional difference decomposition provides critical temporal gradient representations. Similarly, removing the Gated Temporal Convolution (w/o Gated Temporal Conv) reduces accuracy to $54.03\%$, highlighting the importance of channel-wise temporal modeling. In the inter-subject setting, removing Bidirectional Delta drops accuracy by $12.18\%$ (from $55.55\%$ to $43.37\%$), while removing Gated Temporal Conv drops accuracy to $45.83\%$.

On the balanced SADT 2022 dataset (Table~\ref{tab:combined_ablation}), the full DeltaGateNet achieves $96.81\%$ intra-subject and $83.21\%$ inter-subject accuracy. Ablating the Bidirectional Delta module reduces performance to $92.17\%$ and $79.51\%$, respectively, while ablating the Gated Temporal Convolution reduces performance to $87.71\%$ and $75.03\%$.

On the unbalanced SADT 2952 dataset (Table~\ref{tab:combined_ablation}), DeltaGateNet attains $96.84\%$ intra-subject accuracy and $84.49\%$ inter-subject accuracy. In comparison, the configuration w/o Bidirectional Delta achieves $93.30\%$ intra-subject and $77.54\%$ inter-subject accuracy, whereas w/o Gated Temporal Conv achieves $87.07\%$ intra-subject and $81.05\%$ inter-subject accuracy.

These ablation results consistently demonstrate that both the Bidirectional Delta and Gated Temporal Convolution modules are indispensable for high-accuracy vigilance estimation, and their combination enables superior representation learning across diverse subject distributions. To further inspect the structural sensitivities of DeltaGateNet, extensive hyperparameter tuning experiments on temporal convolution kernel sizes $K \in \{1, 3, 5, 7\}$ and differential step sizes $S \in \{1, 2, 3\}$ across both evaluation paradigms are presented in Appendix~\ref{sec:appendix_param_tuning}.

\subsubsection{Influence of Bidirectional Delta}
Following our discussion of individual modules, we examine how the Bidirectional Delta module affects the model. We conducted control experiments across three settings: Signed Delta, Absolute Delta, and Bidirectional Delta. Our results in Table~\ref{tab:ablation_delta} demonstrate that Bidirectional Delta is the best choice. Intuitively, Absolute Delta and Signed Delta both omit information. Signed Delta does not separate the signals which can create anti-correlated signals that are picked up by later convolution modules. In contrast, Absolute Delta treats all negative differentials as positive differentials, omitting directional information.

\subsection{Supplementary Experiments}

\subsubsection{Frequency-Band Analysis of Bidirectional Delta Features}
\begin{figure}[!htbp]
    \centering
    \includegraphics[width=\columnwidth]{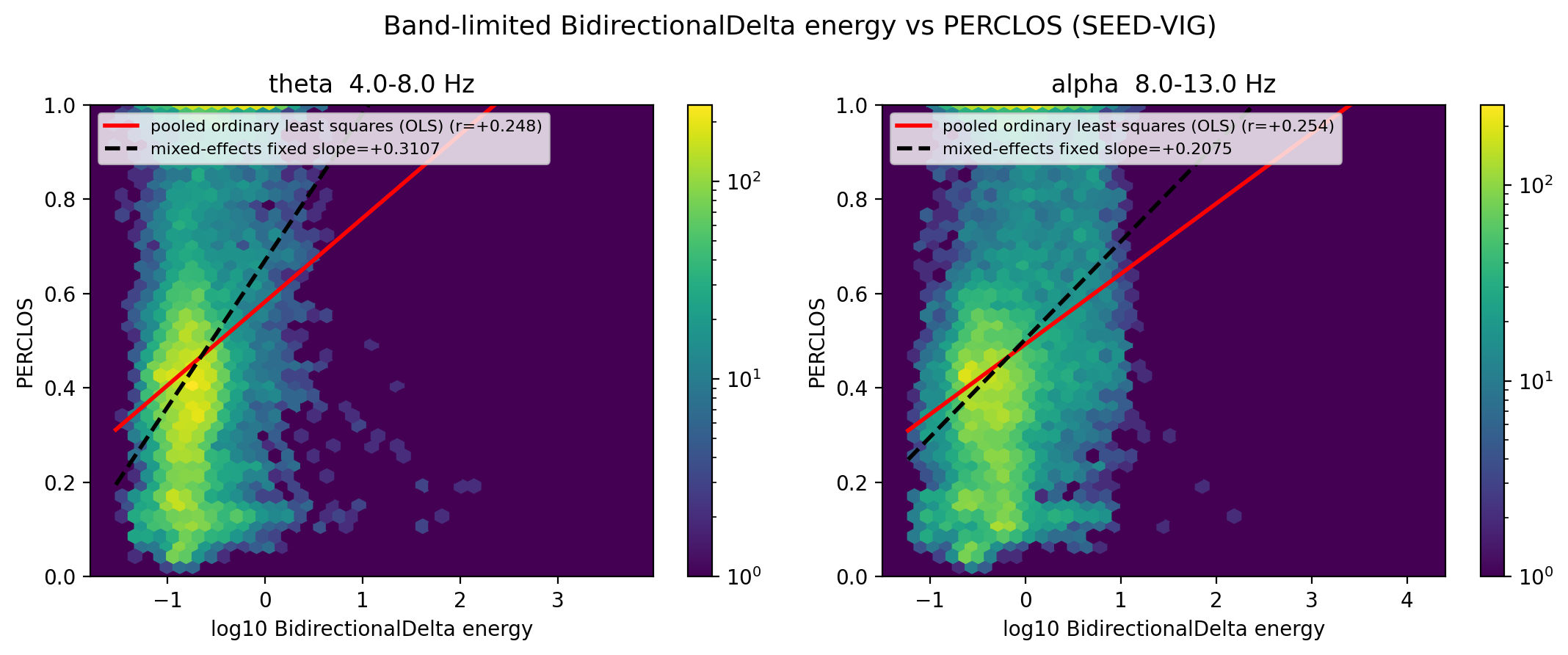}
    \caption{
        Association between band-limited Bidirectional Delta energy and 
        PERCLOS on SEED-VIG. (\textbf{a}--\textbf{b}) Hexbin scatter plots 
        showing $\log_{10}$-transformed Bidirectional Delta energy versus 
        PERCLOS for the theta ($4$--$8$\,Hz) and alpha ($8$--$13$\,Hz) bands. 
        The pooled ordinary least squares (OLS) fit (solid) and the 
        mixed-effects fixed-effect fit (dashed) are overlaid ($N = 20{,}355$ samples, 23 subjects).
    }
    \label{fig:band_correlations}
\end{figure}
To examine the physiological characteristics captured by the Bidirectional Delta representations, we evaluated the association between band-limited Bidirectional Delta energy and continuous PERCLOS on SEED-VIG ($N = 20{,}355$ epochs, 23 subjects), as shown in Fig.~\ref{fig:band_correlations}. A mixed-effects regression with random subject intercepts revealed a robust positive within-subject association between PERCLOS and both theta-band and alpha-band energy.

This association reflects a within-subject temporal process rather than a subject-level trait: the within-subject correlations were positive for theta ($r = +0.380$) and alpha ($r = +0.331$), whereas the between-subject correlations were near zero and non-significant. The effect was highly consistent across individuals (alpha correlated positively with PERCLOS in all 23 subjects and theta in 19 of 23), but its magnitude was modest ($R^2 \approx 6$--$7\%$), indicating that any single band accounts for only a small share of drowsiness-related variance. This suggests that the Bidirectional Delta module captures a broad low-frequency (theta--alpha) slowing associated with increasing drowsiness. Full statistics are tabulated in Appendix~\ref{sec:appendix_freq_band}.

\subsubsection{Per-Class Performance Analysis}
\begin{figure}[htbp]
    \centering
    \includegraphics[width=\columnwidth]{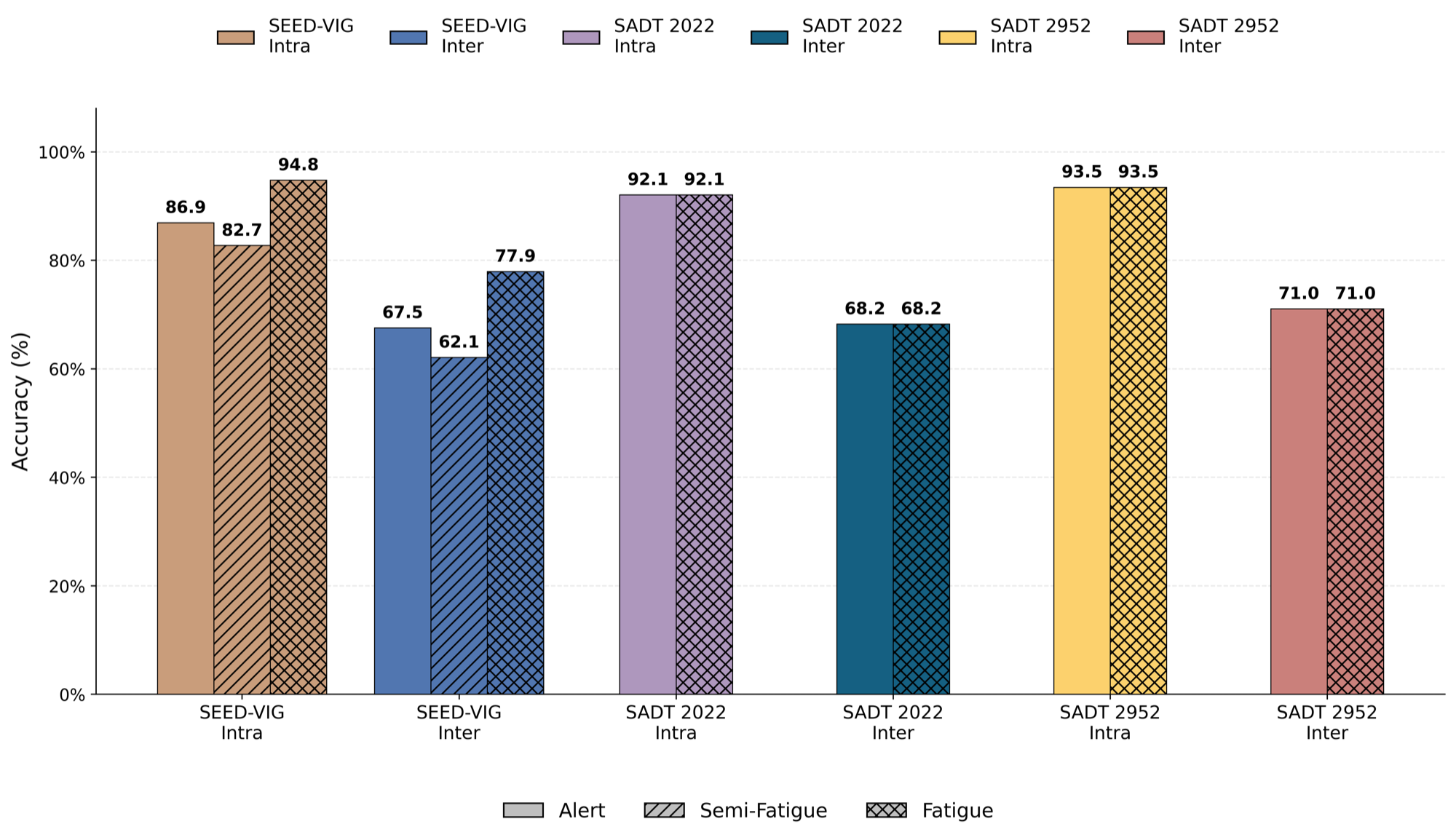}
    \caption{Per-class accuracy of DeltaGateNet across the six (dataset, evaluation setting) groups. Each group is color-coded; hatching distinguishes the vigilance classes (pure: Alert, diagonal stripes: Semi-Fatigue, crossed: Fatigue/Drowsy).}
    \label{fig:class_accuracy}
\end{figure}
\begin{figure}[htbp]
    \centering
    \includegraphics[width=\columnwidth]{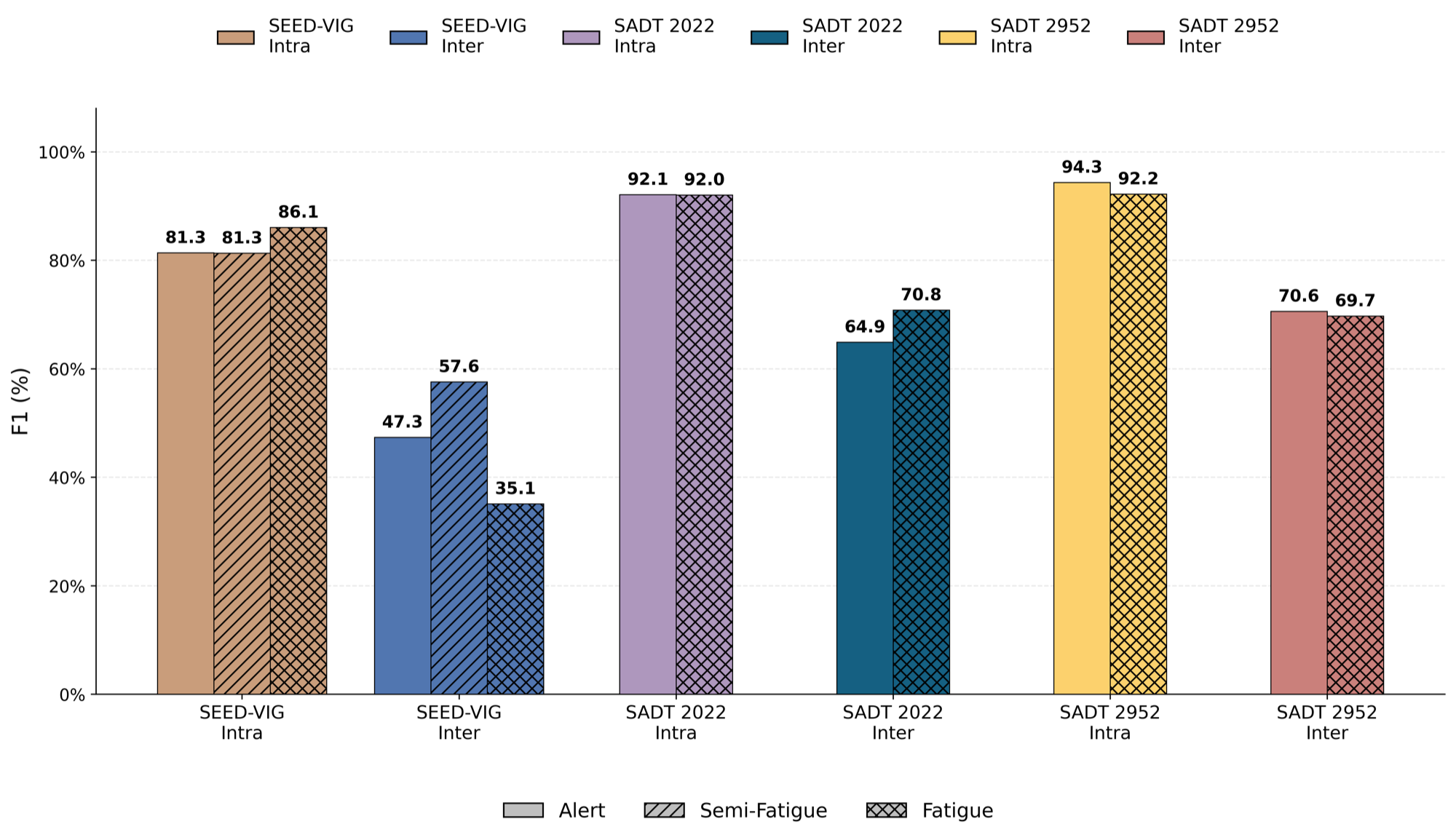}
    \caption{Per-class F1 of DeltaGateNet across the six (dataset, evaluation setting) groups, using the same color and hatching encoding as Fig.~\ref{fig:class_accuracy}.}
    \label{fig:class_f1}
\end{figure}

We evaluate class-wise behavior using two metrics, reported in Fig.~\ref{fig:class_accuracy} (accuracy) and Fig.~\ref{fig:class_f1} (F1). Per-class accuracy for class $c$ is the one-versus-rest accuracy $(\mathrm{TP}_c + \mathrm{TN}_c)/N$. Per-class F1 is the harmonic mean of per-class precision and recall. These metrics complement the overall Accuracy and Macro-F1 in Tables~\ref{tab:intra_subject_results} and~\ref{tab:inter_subject_results} by showing which vigilance states are recovered or confused. On the three-class SEED-VIG dataset, the Fatigue class is the most reliably detected intra-subject, attaining $94.77\%$ accuracy and $86.05\%$ F1, whereas the intermediate Semi-Fatigue class is the most challenging, with an intra-subject accuracy of $82.73\%$. This is consistent with the inherent ambiguity of the transitional vigilance state that lies between alertness and full drowsiness. In the inter-subject setting, the Fatigue class degrades the most, with its F1 falling to $35.09\%$, reflecting a pronounced precision--recall trade-off for this minority class under cross-subject distribution shift. On the binary SADT datasets, one-versus-rest accuracy is identical for Alert and Drowsy by construction and equals overall accuracy; we therefore compare the two classes using per-class F1. On SADT 2022, the intra-subject F1 values are nearly symmetric ($92.11\%$ Alert, $92.03\%$ Drowsy), whereas the inter-subject setting shows a modest drowsy-class advantage ($64.88\%$ Alert, $70.79\%$ Drowsy). On SADT 2952, the per-class F1 values remain closely matched in both paradigms ($94.34\%$ / $92.20\%$ intra-subject and $70.56\%$ / $69.70\%$ inter-subject for Alert / Drowsy), indicating that the model does not systematically favor either state under class imbalance.

\begin{figure*}[t]
    \centering
    \includegraphics[width=\textwidth]{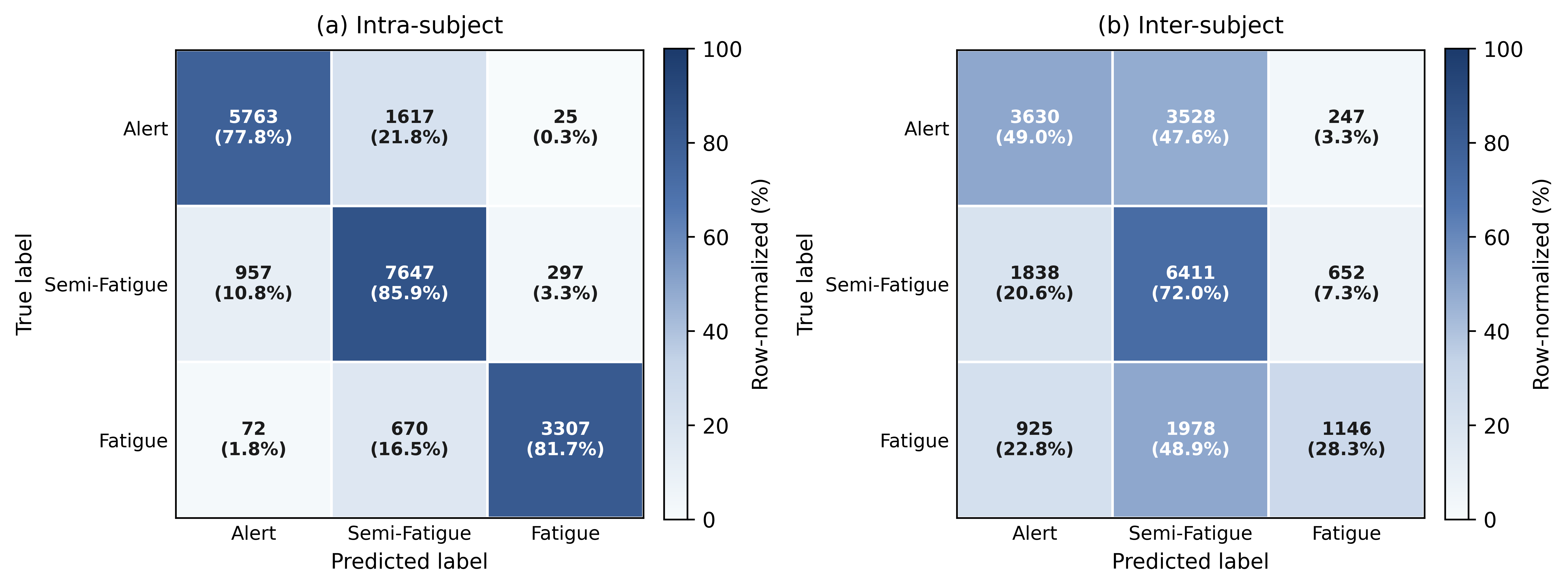}
    \caption{Fold-summed SEED-VIG confusion matrices of DeltaGateNet. Each cell shows the pooled count over five test folds and the corresponding row-normalized percentage. (a) Intra-subject. (b) Inter-subject.}
    \label{fig:seedvig_cm}
\end{figure*}

Fig.~\ref{fig:seedvig_cm} reports fold-summed SEED-VIG confusion matrices (row-normalized percentages with raw counts). Intra-subject errors concentrate on adjacent vigilance states: $21.8\%$ of true Alert epochs are predicted as Semi-Fatigue and $16.5\%$ of true Fatigue epochs as Semi-Fatigue, whereas Alert$\leftrightarrow$Fatigue swaps are rare ($0.3\%$ of true Alert and $1.8\%$ of true Fatigue). Under the inter-subject split the same adjacent-class pattern remains, but Fatigue recall collapses: only $28.3\%$ of true Fatigue epochs are recovered, while $48.9\%$ are labeled Semi-Fatigue and $22.8\%$ are labeled Alert. 

\subsubsection{Inference Latency Analysis}
\input{InferenceTime}

To evaluate the computational efficiency of DeltaGateNet for potential in-vehicle driver monitoring systems, we measured the model forward-pass inference latency across all architectures. As in-vehicle and other edge-device measurements were not available, latency benchmarks were conducted in PyTorch 2.9.0 on CPU with $EEG_{input} \in \mathbb{R}^{1 \times 17 \times 1600}$ corresponding to batch size of 1, electrode channels of 17, and 1600 timesteps.

As reported in Table~\ref{tab:InferenceTime}, DeltaGateNet achieves a forward-pass inference latency of $523.16$~ms per input epoch. This forward pass is distinct from the $8$-second EEG buffer that must be accumulated before each SEED-VIG decision. In an online vehicular monitoring scenario, EEG signals are ingested via a continuous circular buffer, and the model forward pass is executed periodically in a sliding window fashion. Although the forward pass of $523.16$~ms is slightly above the $500$~ms real-time update budget established in automotive monitoring literature \cite{5001,5002}, DeltaGateNet consistently achieves the best classification performance across all datasets and evaluation settings, and remains substantially faster than several stronger baselines. While FCN\_Wang at 422.39~ms may be competitive, but ViT and EEGMamba exceed $900$~ms. Furthermore, DeltaGateNet ingests segmented time-domain EEG directly without requiring multi-stage runtime preprocessing pipelines, such as independent component analysis, adaptive filtering, or wavelet decomposition, which typically add $1$--$2$~seconds of processing latency in conventional workflows \cite{preprocessing}.
\section{Limitations and Discussion}
Despite achieving competitive performance in intra-subject settings, our model exhibits a significant performance degradation in inter-subject scenarios on the more difficult three-class SEED-VIG dataset. 

Furthermore, while our evaluation spans the public driving-fatigue EEG benchmarks that are available at comparable scale, all of these recordings were collected in controlled, simulated environments. This introduces a critical generalization gap, as real-world driving involves continuous motion artifacts, variable lighting, and extreme EEG signal noise that our model has not yet encountered. 

To address these shortcomings, our immediate future work will focus on integrating unsupervised domain adaptation techniques specifically tailored for EEG to align subject-specific feature spaces, and validating our approach on a naturalistic driving dataset currently being collected in our lab. Additionally, we plan to explore lightweight model pruning to minimize the inference latency required for real-time vehicular warning systems.
\section{Conclusion}
In this paper, a bidirectional temporal dynamics modeling framework with gated temporal encoding is proposed for EEG-based driving fatigue recognition. The architecture is designed to address key challenges in intelligent transportation systems (ITS), where continuous physiological monitoring serves as a valuable feedback mechanism for driver safety. Firstly, we introduce a Bidirectional Delta module to decompose first-order temporal differences into positive and negative components, explicitly modeling asymmetric neural activation and suppression patterns associated with vigilance degradation. Secondly, we design a Gated Temporal Convolution module that learns channel-wise temporal dependencies for each EEG channel using temporal convolutions and residual learning, preserving spatial specificity while enhancing temporal representation robustness. Thirdly, we present a complete framework that effectively integrates bidirectional temporal dynamics with channel-wise temporal encoding, improving both feature robustness and cross-subject generalization. By operating directly on segmented time-domain EEG signals without requiring multi-stage runtime filtering or manual artifact cleaning, DeltaGateNet significantly reduces computational complexity. Extensive experiments conducted under both intra-subject and strict subject-disjoint inter-subject evaluation settings on the SADT-2022, SADT 2952, and SEED-VIG datasets demonstrate that DeltaGateNet consistently outperforms existing baseline architectures. While multi-class cross-subject recognition remains an active challenge across the field, the computational efficiency and competitive accuracy of DeltaGateNet highlight its potential for integration into in-vehicle driver monitoring systems. In future work, we will explore subject-calibration and domain adaptation techniques to further improve cross-subject performance.

\clearpage
\appendices

\section{Hyperparameter Sensitivity Analysis}
\label{sec:appendix_param_tuning}
To inspect the structural hyperparameter sensitivities of DeltaGateNet, we conducted systematic experiments evaluating the temporal convolution kernel size $K \in \{1, 3, 5, 7\}$ in the Gated Temporal Convolution module and the discrete differential step size $S \in \{1, 2, 3\}$ ($\Delta x(t) = x(t) - x(t-S)$) in the Bidirectional Delta module. The results on SEED-VIG, SADT 2022, and SADT 2952 under both intra-subject and inter-subject evaluation paradigms are reported in Table~\ref{tab:combined_kernel_size} and Table~\ref{tab:combined_step_size}.

\input{Parameter_Tuning}

\section{Class Distribution Statistics}
\label{sec:appendix_class_distribution}
Table~\ref{tab:class_distribution} reports the detailed per-class sample counts and percentages for the three benchmark datasets. SEED-VIG uses a three-class labeling derived from PERCLOS, whereas both SADT releases use a binary alert/drowsy labeling.

\begin{table}[H]
\centering
\caption{Per-class sample counts and percentages of the three benchmark datasets.}
\label{tab:class_distribution}
\resizebox{\columnwidth}{!}{%
\begin{tabular}{@{}lccccc@{}}
\toprule
\textbf{Dataset} & \textbf{Total} & \textbf{Alert} & \textbf{Semi-Fatigue} & \textbf{Fatigue} & \textbf{Drowsy} \\
\midrule
SEED-VIG & 20{,}355 & 7{,}405 (36.38\%) & 8{,}901 (43.73\%) & 4{,}049 (19.89\%) & --- \\
SADT 2022 & 2{,}022 & 1{,}011 (50.00\%) & --- & --- & 1{,}011 (50.00\%) \\
SADT 2952 & 2{,}952 & 1{,}731 (58.64\%) & --- & --- & 1{,}221 (41.36\%) \\
\bottomrule
\end{tabular}%
}
\end{table}

\section{Frequency-Band Analysis: Full Statistics}
\label{sec:appendix_freq_band}
This appendix reports the per-band statistics for the association between band-limited Bidirectional Delta energy and continuous PERCLOS on SEED-VIG. Theta- and alpha-band energies show a consistent positive within-subject association, whereas delta-band energy shows no consistent association and beta-band energy is subject-wise inconsistent.

\begin{table}[H]
\centering
\caption{Mixed-effects model $PERCLOS \sim band\_energy + (1|subject)$ for all four bands.}
\label{tab:freq_lmm}
\resizebox{\columnwidth}{!}{%
\begin{tabular}{@{}lccccc@{}}
\toprule
\textbf{Band} & \textbf{Fixed slope} & \textbf{SE} & \textbf{$z$} & \textbf{$p$} & \textbf{95\% CI} \\
\midrule
Delta ($0.5$--$4$\,Hz) & $+0.086$ & $0.0044$ & $+19.8$ & $<0.001$ & $[+0.078,+0.095]$ \\
Theta ($4$--$8$\,Hz) & $+0.311$ & $0.0053$ & $+58.5$ & $<0.001$ & $[+0.300,+0.321]$ \\
Alpha ($8$--$13$\,Hz) & $+0.208$ & $0.0041$ & $+50.1$ & $<0.001$ & $[+0.199,+0.216]$ \\
Beta ($13$--$30$\,Hz) & $-0.075$ & $0.0058$ & $-13.0$ & $<0.001$ & $[-0.086,-0.064]$ \\
\bottomrule
\end{tabular}%
}
\end{table}

\begin{table}[H]
\centering
\caption{Subject-wise Pearson correlations between band energy and PERCLOS}
\label{tab:freq_subjectwise}
\resizebox{\columnwidth}{!}{%
\begin{tabular}{@{}lccccc@{}}
\toprule
\textbf{Band} & \textbf{Mean $r$} & \textbf{Median $r$} & \textbf{Range} & \textbf{Pos./Neg.} & \textbf{Wilcoxon $p$} \\
\midrule
Delta ($0.5$--$4$\,Hz) & $+0.097$ & $-0.002$ & $[-0.221,+0.620]$ & $10/11$ & $0.30$ \\
Theta ($4$--$8$\,Hz) & $+0.350$ & $+0.350$ & $[-0.227,+0.820]$ & $19/2$ & $2.4\times10^{-5}$ \\
Alpha ($8$--$13$\,Hz) & $+0.318$ & $+0.280$ & $[+0.043,+0.786]$ & $21/0$ & $9.5\times10^{-7}$ \\
Beta ($13$--$30$\,Hz) & $-0.063$ & $-0.094$ & $[-0.424,+0.354]$ & $8/13$ & $0.17$ \\
\bottomrule
\end{tabular}%
}
\end{table}

\begin{table}[H]
\centering
\caption{Pairwise band specificity. Paired permutation and null-centered bootstrap $t$-tests on the per-subject Fisher-$z$ correlations, averaged over $7$ runs ($\text{seed} = 2026$--$2032$); values are mean $\pm$ std.}
\label{tab:freq_specificity}
\resizebox{\columnwidth}{!}{%
\begin{tabular}{@{}lcc@{}}
\toprule
\textbf{Band pair} & \textbf{Permutation $p$} & \textbf{Bootstrap $t$ $p$} \\
\midrule
Delta vs.\ Theta & $0.0002 \pm 0.0001$ & $0.0003 \pm 0.0001$ \\
Delta vs.\ Alpha & $0.0133 \pm 0.0010$ & $0.0129 \pm 0.0012$ \\
Delta vs.\ Beta & $0.0476 \pm 0.0030$ & $0.0651 \pm 0.0033$ \\
Theta vs.\ Alpha & $0.6092 \pm 0.0114$ & $0.6037 \pm 0.0069$ \\
Theta vs.\ Beta & $0.0003 \pm 0.0001$ & $0.0003 \pm 0.0001$ \\
Alpha vs.\ Beta & $0.0002 \pm 0.0000$ & $0.0003 \pm 0.0001$ \\
\bottomrule
\end{tabular}%
}
\end{table}

\begin{table}[H]
\centering
\caption{Within- versus between-subject decomposition of the band energy--PERCLOS correlation.}
\label{tab:freq_withinbetween}
\resizebox{\columnwidth}{!}{%
\begin{tabular}{@{}lcc@{}}
\toprule
\textbf{Band} & \textbf{Within-subject $r$} & \textbf{Between-subject $r$} \\
\midrule
Delta ($0.5$--$4$\,Hz) & $+0.137$ ($p<0.001$) & $-0.130$ (n.s.) \\
Theta ($4$--$8$\,Hz) & $+0.380$ ($p<0.001$) & $-0.069$ (n.s.) \\
Alpha ($8$--$13$\,Hz) & $+0.331$ ($p<0.001$) & $+0.014$ (n.s.) \\
Beta ($13$--$30$\,Hz) & $-0.090$ ($p<0.001$) & $-0.296$ (n.s.) \\
\bottomrule
\end{tabular}%
}
\end{table}

\end{document}

%% file: Intrasubject.tex
\begin{table*}[t]
\centering
\caption{Intra-Subject Evaluation Results}
\label{tab:intra_subject_results}
\begin{threeparttable}
\resizebox{\textwidth}{!}{%
\begin{tabular}{@{}lcccccccccccc@{}}
\toprule
\multirow{2}{*}{\textbf{Model}} & \multicolumn{4}{c}{\textbf{SEED-VIG}} & \multicolumn{4}{c}{\textbf{SADT 2022}} & \multicolumn{4}{c}{\textbf{SADT 2952}} \\
\cmidrule(lr){2-5} \cmidrule(lr){6-9} \cmidrule(lr){10-13}
& \textbf{Acc} $\uparrow$ & \textbf{Pre} $\uparrow$ & \textbf{Rec} $\uparrow$ & \textbf{F1} $\uparrow$ & \textbf{Acc} $\uparrow$ & \textbf{Pre} $\uparrow$ & \textbf{Rec} $\uparrow$ & \textbf{F1} $\uparrow$ & \textbf{Acc} $\uparrow$ & \textbf{Pre} $\uparrow$ & \textbf{Rec} $\uparrow$ & \textbf{F1} $\uparrow$ \\
\midrule
\rowcolor{catBlue}
\multicolumn{13}{c}{\rule[-0.5em]{0pt}{1.7em}\textbf{Time Series Classification Models}} \\
\addlinespace[0.4em]
FCN\_Wang \cite{Resnet_Wang} & $68.03 \pm 1.19$ & $70.07 \pm 1.62$ & $68.38 \pm 0.75$ & $68.50 \pm 0.97$ & $\underline{96.33 \pm 1.65}$ & $\underline{96.32 \pm 1.63}$ & $\underline{96.37 \pm 1.61}$ & $\underline{96.32 \pm 1.64}$ & $\underline{96.66 \pm 2.06}$ & $\underline{96.39 \pm 2.20}$ & $\underline{96.52 \pm 2.29}$ & $\underline{96.45 \pm 2.24}$ \\
InceptionTime \cite{InceptionTime} & $77.41 \pm 0.63$ & $79.13 \pm 0.61$ & $77.85 \pm 1.07$ & $78.30 \pm 0.79$ & $94.35 \pm 4.98$ & $94.32 \pm 4.95$ & $94.39 \pm 4.98$ & $94.34 \pm 4.97$ & $95.37 \pm 3.13$ & $95.30 \pm 3.32$ & $95.17 \pm 3.06$ & $95.22 \pm 3.18$ \\
Resnet\_Wang \cite{Resnet_Wang} & $77.64 \pm 0.68$ & $78.92 \pm 0.79$ & $78.44 \pm 0.82$ & $78.61 \pm 0.62$ & $93.32 \pm 5.28$ & $93.70 \pm 4.74$ & $93.39 \pm 5.08$ & $93.30 \pm 5.31$ & $94.07 \pm 3.32$ & $94.50 \pm 3.14$ & $93.28 \pm 4.01$ & $93.74 \pm 3.72$ \\
\addlinespace[0.7em]
\rowcolor{catPurple}
\multicolumn{13}{c}{\rule[-0.5em]{0pt}{1.7em}\textbf{Deep Learning CNN and Transformer-Based Models}} \\
\addlinespace[0.4em]
XResnet \cite{XResnet101} & $63.71 \pm 2.42$ & $67.33 \pm 2.01$ & $63.20 \pm 3.14$ & $63.54 \pm 2.66$ & $85.42 \pm 8.05$ & $86.20 \pm 7.57$ & $85.31 \pm 8.12$ & $85.24 \pm 8.22$ & $88.84 \pm 7.24$ & $88.60 \pm 7.43$ & $88.25 \pm 7.61$ & $88.37 \pm 7.50$ \\
ViT \cite{ViT} & $71.59 \pm 0.50$ & $73.85 \pm 1.13$ & $71.90 \pm 1.11$ & $72.41 \pm 0.58$ & $86.53 \pm 4.92$ & $86.85 \pm 4.96$ & $86.57 \pm 4.89$ & $86.50 \pm 4.92$ & $88.42 \pm 5.30$ & $88.57 \pm 5.45$ & $87.65 \pm 5.41$ & $87.95 \pm 5.38$ \\
\addlinespace[0.7em]
\rowcolor{catGreen}
\multicolumn{13}{c}{\rule[-0.5em]{0pt}{1.7em}\textbf{EEG-Specific Models}} \\
\addlinespace[0.4em]
EEGNet \cite{EEGNet} & $73.10 \pm 1.94$ & $75.44 \pm 1.23$ & $73.94 \pm 1.46$ & $73.97 \pm 1.88$ & $76.78 \pm 7.41$ & $81.34 \pm 4.24$ & $76.99 \pm 6.84$ & $75.69 \pm 8.46$ & $88.35 \pm 2.78$ & $89.01 \pm 2.57$ & $87.01 \pm 3.21$ & $87.68 \pm 3.07$ \\
TSception \cite{TSception} & \underline{$79.80 \pm 0.98$} & \underline{$81.26 \pm 0.80$} & \underline{$80.29 \pm 1.11$} & \underline{$80.68 \pm 0.95$} & $82.51 \pm 2.99$ & $85.77 \pm 2.62$ & $82.46 \pm 3.24$ & $82.04 \pm 3.27$ & $73.07 \pm 5.13$ & $79.60 \pm 5.02$ & $67.54 \pm 5.16$ & $67.01 \pm 6.56$ \\
GAT-CNN \cite{GATCNN} & $66.52 \pm 1.07$ & $68.32 \pm 3.23$ & $66.68 \pm 1.48$ & $66.58 \pm 1.29$ & $71.30 \pm 1.95$ & $79.63 \pm 1.99$ & $57.31 \pm 5.03$ & $66.49 \pm 3.31$ & $71.61 \pm 1.78$ & $78.05 \pm 2.96$ & $44.01 \pm 4.26$ & $56.06 \pm 2.75$ \\
\addlinespace[0.7em]
\rowcolor{catOrange}
\multicolumn{13}{c}{\rule[-0.5em]{0pt}{1.7em}\textbf{EEG Foundation Models}} \\
\addlinespace[0.4em]
LaBraM \cite{LaBraM} & $76.41 \pm 0.98$ & $77.90 \pm 1.08$ & $76.83 \pm 1.12$ & $77.23 \pm 0.99$ & $82.54 \pm 0.64$ & $82.64 \pm 0.59$ & $82.56 \pm 0.59$ & $82.52 \pm 0.63$ & $83.05 \pm 1.77$ & $82.51 \pm 1.59$ & $82.60 \pm 1.54$ & $82.54 \pm 1.57$ \\
EEGMamba \cite{EEGMamba} & $62.79 \pm 1.37$ & $64.24 \pm 2.24$ & $61.36 \pm 0.89$ & $61.93 \pm 1.44$ & $65.59 \pm 1.71$ & $66.50 \pm 1.65$ & $65.48 \pm 1.72$ & $64.99 \pm 2.02$ & $67.16 \pm 3.34$ & $66.19 \pm 3.72$ & $65.23 \pm 2.89$ & $65.38 \pm 3.03$ \\
\addlinespace[0.4em]
\midrule
\textbf{DeltaGateNet (Ours)} & $\mathbf{81.89 \pm 0.66}$ & $\mathbf{84.37 \pm 0.52}$ & $\mathbf{81.32 \pm 0.64}$ & $\mathbf{82.55 \pm 0.58}$ & $\mathbf{96.81 \pm 2.58}$ & $\mathbf{96.98 \pm 2.34}$ & $\mathbf{96.80 \pm 2.60}$ & $\mathbf{96.81 \pm 2.59}$ & $\mathbf{96.84 \pm 1.43}$ & $\mathbf{96.73 \pm 1.41}$ & $\mathbf{96.81 \pm 1.53}$ & $\mathbf{96.77 \pm 1.47}$ \\
\bottomrule
\end{tabular}%
}
\end{threeparttable}
\end{table*}

%% file: Intersubject.tex
\begin{table*}[t]
\centering
\caption{Inter-Subject Evaluation Results}
\label{tab:inter_subject_results}
\begin{threeparttable}
\resizebox{\textwidth}{!}{%
\begin{tabular}{@{}lcccccccccccc@{}}
\toprule
\multirow{2}{*}{\textbf{Model}} & \multicolumn{4}{c}{\textbf{SEED-VIG}} & \multicolumn{4}{c}{\textbf{SADT 2022}} & \multicolumn{4}{c}{\textbf{SADT 2952}} \\
\cmidrule(lr){2-5} \cmidrule(lr){6-9} \cmidrule(lr){10-13}
& \textbf{Acc} $\uparrow$ & \textbf{Pre} $\uparrow$ & \textbf{Rec} $\uparrow$ & \textbf{F1} $\uparrow$ & \textbf{Acc} $\uparrow$ & \textbf{Pre} $\uparrow$ & \textbf{Rec} $\uparrow$ & \textbf{F1} $\uparrow$ & \textbf{Acc} $\uparrow$ & \textbf{Pre} $\uparrow$ & \textbf{Rec} $\uparrow$ & \textbf{F1} $\uparrow$ \\
\midrule
\rowcolor{catBlue}
\multicolumn{13}{c}{\rule[-0.5em]{0pt}{1.7em}\textbf{Time Series Classification Models}} \\
\addlinespace[0.4em]
FCN\_Wang \cite{Resnet_Wang} & $\underline{48.91 \pm 8.05}$ & $\underline{50.10 \pm 10.98}$ & $\underline{48.66 \pm 9.10}$ & $\underline{46.16 \pm 8.63}$ & $73.93 \pm 10.29$ & $75.33 \pm 10.04$ & $73.93 \pm 10.29$ & $73.48 \pm 10.52$ & $\underline{82.99 \pm 11.61}$ & $\underline{82.46 \pm 9.80}$ & $\underline{83.21 \pm 12.18}$ & $\underline{81.14 \pm 12.70}$ \\
InceptionTime \cite{InceptionTime} & $47.05 \pm 5.38$ & $48.27 \pm 5.26$ & $45.98 \pm 8.42$ & $43.96 \pm 8.73$ & $\underline{80.12 \pm 9.11}$ & $\underline{81.37 \pm 8.28}$ & $\underline{80.12 \pm 9.11}$ & $\underline{79.70 \pm 9.71}$ & $76.03 \pm 4.12$ & $77.32 \pm 2.30$ & $75.90 \pm 3.46$ & $74.28 \pm 3.55$ \\
Resnet\_Wang \cite{Resnet_Wang} & $45.61 \pm 5.66$ & $42.53 \pm 9.46$ & $44.66 \pm 6.32$ & $41.35 \pm 8.49$ & $72.67 \pm 7.74$ & $74.90 \pm 6.82$ & $72.67 \pm 7.74$ & $71.72 \pm 8.61$ & $72.47 \pm 12.81$ & $74.04 \pm 12.42$ & $71.65 \pm 12.10$ & $71.00 \pm 13.66$ \\
\addlinespace[0.7em]
\rowcolor{catPurple}
\multicolumn{13}{c}{\rule[-0.5em]{0pt}{1.7em}\textbf{Deep Learning CNN and Transformer-Based Models}} \\
\addlinespace[0.4em]
XResnet \cite{XResnet101} & $45.76 \pm 4.28$ & $48.32 \pm 5.42$ & $44.07 \pm 4.78$ & $42.09 \pm 7.00$ & $64.07 \pm 9.99$ & $72.52 \pm 7.67$ & $64.07 \pm 9.99$ & $58.91 \pm 13.75$ & $66.11 \pm 6.25$ & $68.18 \pm 7.37$ & $61.33 \pm 6.34$ & $58.74 \pm 8.86$ \\
ViT \cite{ViT} & $46.06 \pm 4.62$ & $47.75 \pm 8.24$ & $43.00 \pm 4.90$ & $40.40 \pm 4.27$ & $75.54 \pm 6.38$ & $77.03 \pm 6.25$ & $75.54 \pm 6.38$ & $75.13 \pm 6.65$ & $78.58 \pm 4.60$ & $77.46 \pm 4.64$ & $79.37 \pm 6.28$ & $77.37 \pm 5.04$ \\
\addlinespace[0.7em]
\rowcolor{catGreen}
\multicolumn{13}{c}{\rule[-0.5em]{0pt}{1.7em}\textbf{EEG-Specific Models}} \\
\addlinespace[0.4em]
EEGNet \cite{EEGNet} & $44.64 \pm 5.99$ & $46.56 \pm 8.93$ & $43.21 \pm 4.96$ & $39.52 \pm 4.73$ & $68.99 \pm 4.84$ & $74.77 \pm 6.49$ & $68.99 \pm 4.84$ & $67.13 \pm 5.10$ & $77.01 \pm 11.27$ & $79.96 \pm 11.07$ & $76.00 \pm 10.34$ & $75.22 \pm 12.49$ \\
TSception \cite{TSception} & $42.19 \pm 3.90$ & $42.98 \pm 7.89$ & $41.91 \pm 5.61$ & $38.64 \pm 7.80$ & $73.71 \pm 4.68$ & $76.89 \pm 3.68$ & $73.71 \pm 4.68$ & $72.79 \pm 5.47$ & $58.25 \pm 7.80$ & $66.38 \pm 2.37$ & $62.30 \pm 2.66$ & $55.56 \pm 6.15$ \\
GAT-CNN \cite{GATCNN} & $41.26 \pm 5.99$ & $37.96 \pm 6.30$ & $41.74 \pm 3.84$ & $36.22 \pm 6.17$ & $64.10 \pm 6.80$ & $79.42 \pm 12.05$ & $36.92 \pm 10.17$ & $50.15 \pm 11.99$ & $60.54 \pm 12.75$ & $57.80 \pm 19.06$ & $21.57 \pm 17.43$ & $29.06 \pm 21.81$ \\
\addlinespace[0.7em]
\rowcolor{catOrange}
\multicolumn{13}{c}{\rule[-0.5em]{0pt}{1.7em}\textbf{EEG Foundation Models}} \\
\addlinespace[0.4em]
LaBraM \cite{LaBraM} & $43.20 \pm 7.67$ & $39.86 \pm 10.14$ & $45.27 \pm 5.81$ & $37.20 \pm 8.93$ & $70.58 \pm 6.95$ & $72.04 \pm 7.73$ & $70.58 \pm 6.95$ & $70.15 \pm 6.99$ & $60.04 \pm 8.43$ & $54.65 \pm 15.02$ & $58.83 \pm 8.45$ & $54.87 \pm 12.07$ \\
EEGMamba \cite{EEGMamba} & $47.87 \pm 4.21$ & $46.61 \pm 6.31$ & $45.03 \pm 5.81$ & $43.21 \pm 5.78$ & $61.36 \pm 3.19$ & $62.09 \pm 3.26$ & $61.36 \pm 3.19$ & $60.73 \pm 3.46$ & $54.70 \pm 8.60$ & $55.68 \pm 4.41$ & $54.62 \pm 3.67$ & $51.83 \pm 5.86$ \\
\addlinespace[0.4em]
\midrule
\textbf{DeltaGateNet (Ours)} & $\mathbf{55.55 \pm 3.96}$ & $\mathbf{54.20 \pm 12.81}$ & $\mathbf{48.32 \pm 7.60}$ & $\mathbf{47.32 \pm 8.21}$ & $\mathbf{83.21 \pm 6.25}$ & $\mathbf{84.18 \pm 5.50}$ & $\mathbf{83.21 \pm 6.25}$ & $\mathbf{83.03 \pm 6.41}$ & $\mathbf{84.49 \pm 7.36}$ & $\mathbf{85.66 \pm 6.83}$ & $\mathbf{83.90 \pm 5.85}$ & $\mathbf{83.92 \pm 6.93}$ \\
\bottomrule
\end{tabular}%
}
\end{threeparttable}
\end{table*}

%% file: Ablation.tex
\begin{table*}[t]
\centering
\caption{Component Ablation Evaluation Results}
\label{tab:combined_ablation}
\begin{threeparttable}
\resizebox{\textwidth}{!}{%
\begin{tabular}{@{}lcccccccccccc@{}}
\toprule
\multirow{2}{*}{\textbf{Model Configuration}} & \multicolumn{4}{c}{\textbf{SEED-VIG}} & \multicolumn{4}{c}{\textbf{SADT 2022}} & \multicolumn{4}{c}{\textbf{SADT 2952}} \\
\cmidrule(lr){2-5} \cmidrule(lr){6-9} \cmidrule(lr){10-13}
& \textbf{Acc} $\uparrow$ & \textbf{Prec} $\uparrow$ & \textbf{Rec} $\uparrow$ & \textbf{F1} $\uparrow$ & \textbf{Acc} $\uparrow$ & \textbf{Prec} $\uparrow$ & \textbf{Rec} $\uparrow$ & \textbf{F1} $\uparrow$ & \textbf{Acc} $\uparrow$ & \textbf{Prec} $\uparrow$ & \textbf{Rec} $\uparrow$ & \textbf{F1} $\uparrow$ \\
\midrule
\rowcolor{tablecategory}
\multicolumn{13}{c}{\rule[-0.5em]{0pt}{1.7em}\textbf{Intra-Subject Evaluation}} \\
\addlinespace[0.25em]
w/o Both Modules (MLP only) & $43.72\!\pm\!0.74$ & $30.87\!\pm\!2.20$ & $33.35\!\pm\!0.03$ & $20.46\!\pm\!0.26$ & $72.53\!\pm\!11.83$ & $72.53\!\pm\!11.98$ & $72.52\!\pm\!11.97$ & $72.44\!\pm\!11.97$ & $74.30\!\pm\!8.73$ & $74.71\!\pm\!10.33$ & $71.36\!\pm\!9.14$ & $71.68\!\pm\!10.09$ \\
w/o Gated Temporal Conv & $54.03\!\pm\!1.69$ & $64.69\!\pm\!2.99$ & $45.59\!\pm\!1.37$ & $43.13\!\pm\!1.01$ & $87.71\!\pm\!13.04$ & $87.72\!\pm\!13.07$ & $87.66\!\pm\!13.15$ & $87.64\!\pm\!13.17$ & $87.07\!\pm\!10.12$ & $86.17\!\pm\!11.08$ & $86.18\!\pm\!11.26$ & $86.14\!\pm\!11.17$ \\
w/o Bidirectional Delta & $\underline{71.01\!\pm\!1.12}$ & $\underline{71.87\!\pm\!0.96}$ & $\underline{71.81\!\pm\!1.26}$ & $\underline{71.67\!\pm\!0.96}$ & $\underline{92.17\!\pm\!2.94}$ & $\underline{92.28\!\pm\!3.01}$ & $\underline{92.17\!\pm\!2.99}$ & $\underline{92.15\!\pm\!2.95}$ & $\underline{93.30\!\pm\!3.35}$ & $\underline{93.43\!\pm\!2.95}$ & $\underline{92.86\!\pm\!3.80}$ & $\underline{92.99\!\pm\!3.52}$ \\
\midrule
\textbf{DeltaGateNet (Full Model)} & $\mathbf{81.89\!\pm\!0.66}$ & $\mathbf{84.37\!\pm\!0.52}$ & $\mathbf{81.32\!\pm\!0.64}$ & $\mathbf{82.55\!\pm\!0.58}$ & $\mathbf{96.81\!\pm\!2.58}$ & $\mathbf{96.98\!\pm\!2.34}$ & $\mathbf{96.80\!\pm\!2.60}$ & $\mathbf{96.81\!\pm\!2.59}$ & $\mathbf{96.84\!\pm\!1.43}$ & $\mathbf{96.73\!\pm\!1.41}$ & $\mathbf{96.81\!\pm\!1.53}$ & $\mathbf{96.77\!\pm\!1.47}$ \\
\addlinespace[0.55em]
\rowcolor{tablecategory}
\multicolumn{13}{c}{\rule[-0.5em]{0pt}{1.7em}\textbf{Inter-Subject Evaluation}} \\
\addlinespace[0.25em]
w/o Both Modules (MLP only) & $43.86\!\pm\!7.98$ & $24.96\!\pm\!8.41$ & $33.37\!\pm\!0.87$ & $22.09\!\pm\!3.47$ & $69.73\!\pm\!10.06$ & $71.61\!\pm\!7.16$ & $69.73\!\pm\!10.06$ & $66.62\!\pm\!15.90$ & $70.28\!\pm\!12.88$ & $69.98\!\pm\!14.39$ & $65.25\!\pm\!10.71$ & $65.87\!\pm\!12.16$ \\
w/o Gated Temporal Conv & $\underline{45.83\!\pm\!1.77}$ & $\underline{34.38\!\pm\!4.08}$ & $\underline{37.32\!\pm\!1.39}$ & $\underline{31.62\!\pm\!4.46}$ & $75.03\!\pm\!13.83$ & $71.63\!\pm\!19.22$ & $70.74\!\pm\!18.89$ & $70.35\!\pm\!19.10$ & $\underline{81.05\!\pm\!14.05}$ & $\underline{80.10\!\pm\!14.06}$ & $\underline{78.16\!\pm\!13.20}$ & $\underline{78.78\!\pm\!13.45}$ \\
w/o Bidirectional Delta & $43.37\!\pm\!4.13$ & $43.54\!\pm\!5.69$ & $40.98\!\pm\!3.15$ & $47.32\!\pm\!8.21$ & $\underline{79.51\!\pm\!1.76}$ & $\underline{80.53\!\pm\!1.60}$ & $\underline{79.51\!\pm\!1.76}$ & $\underline{79.34\!\pm\!1.85}$ & $77.54\!\pm\!8.46$ & $78.93\!\pm\!3.65$ & $78.94\!\pm\!5.89$ & $76.55\!\pm\!8.13$ \\
\midrule
\textbf{DeltaGateNet (Full Model)} & $\mathbf{55.55\!\pm\!3.96}$ & $\mathbf{54.20\!\pm\!12.81}$ & $\mathbf{48.32\!\pm\!7.60}$ & $\mathbf{47.32\!\pm\!8.21}$ & $\mathbf{83.21\!\pm\!6.25}$ & $\mathbf{84.18\!\pm\!5.50}$ & $\mathbf{83.21\!\pm\!6.25}$ & $\mathbf{83.03\!\pm\!6.41}$ & $\mathbf{84.49\!\pm\!7.36}$ & $\mathbf{85.66\!\pm\!6.83}$ & $\mathbf{83.90\!\pm\!5.85}$ & $\mathbf{83.92\!\pm\!6.93}$ \\
\bottomrule
\end{tabular}%
}
\end{threeparttable}
\end{table*}

%% file: Ablation_Delta.tex
\begin{table*}[t]
\centering
\caption{Delta Ablation Evaluation Results}
\label{tab:ablation_delta}
\begin{threeparttable}
\resizebox{\textwidth}{!}{%
\begin{tabular}{@{}lcccccccccccc@{}}
\toprule
\multirow{2}{*}{\textbf{Model Configuration}} & \multicolumn{4}{c}{\textbf{SEED-VIG}} & \multicolumn{4}{c}{\textbf{SADT 2022}} & \multicolumn{4}{c}{\textbf{SADT 2952}} \\
\cmidrule(lr){2-5} \cmidrule(lr){6-9} \cmidrule(lr){10-13}
& \textbf{Acc} $\uparrow$ & \textbf{Prec} $\uparrow$ & \textbf{Rec} $\uparrow$ & \textbf{F1} $\uparrow$ & \textbf{Acc} $\uparrow$ & \textbf{Prec} $\uparrow$ & \textbf{Rec} $\uparrow$ & \textbf{F1} $\uparrow$ & \textbf{Acc} $\uparrow$ & \textbf{Prec} $\uparrow$ & \textbf{Rec} $\uparrow$ & \textbf{F1} $\uparrow$ \\
\midrule
\rowcolor{tablecategory}
\multicolumn{13}{c}{\rule[-0.5em]{0pt}{1.7em}\textbf{Intra-Subject Evaluation}} \\
\addlinespace[0.25em]
Signed Delta & $\underline{80.19\!\pm\!1.10}$ & $\underline{82.38\!\pm\!1.13}$ & $\underline{80.01\!\pm\!1.40}$ & $\underline{80.98\!\pm\!1.13}$ & $\underline{91.32\!\pm\!1.80}$ & $\underline{91.50\!\pm\!1.75}$ & $\underline{91.34\!\pm\!1.69}$ & $\underline{91.31\!\pm\!1.79}$ & $\underline{92.67\!\pm\!0.84}$ & $\underline{92.39\!\pm\!0.96}$ & $\underline{92.51\!\pm\!0.94}$ & $\underline{92.43\!\pm\!0.93}$ \\
Absolute Delta & $73.42\!\pm\!2.50$ & $77.82\!\pm\!2.04$ & $71.23\!\pm\!3.78$ & $72.85\!\pm\!2.99$ & $87.91\!\pm\!1.28$ & $87.92\!\pm\!1.30$ & $87.91\!\pm\!1.29$ & $87.90\!\pm\!1.28$ & $88.72\!\pm\!2.94$ & $88.34\!\pm\!3.15$ & $88.60\!\pm\!2.78$ & $88.41\!\pm\!2.98$ \\
\midrule
\textbf{Bidirectional Delta} & $\mathbf{81.89\!\pm\!0.66}$ & $\mathbf{84.37\!\pm\!0.52}$ & $\mathbf{81.32\!\pm\!0.64}$ & $\mathbf{82.55\!\pm\!0.58}$ & $\mathbf{96.81\!\pm\!2.58}$ & $\mathbf{96.98\!\pm\!2.34}$ & $\mathbf{96.80\!\pm\!2.60}$ & $\mathbf{96.81\!\pm\!2.59}$ & $\mathbf{96.84\!\pm\!1.43}$ & $\mathbf{96.73\!\pm\!1.41}$ & $\mathbf{96.81\!\pm\!1.53}$ & $\mathbf{96.77\!\pm\!1.47}$ \\
\addlinespace[0.55em]
\rowcolor{tablecategory}
\multicolumn{13}{c}{\rule[-0.5em]{0pt}{1.7em}\textbf{Inter-Subject Evaluation}} \\
\addlinespace[0.25em]
Signed Delta & $41.78\!\pm\!4.84$ & $41.17\!\pm\!5.71$ & $\underline{38.47\!\pm\!4.89}$ & $\underline{37.70\!\pm\!6.04}$ & $\underline{70.42\!\pm\!6.57}$ & $\underline{71.48\!\pm\!6.01}$ & $\underline{70.42\!\pm\!6.57}$ & $\underline{69.83\!\pm\!7.22}$ & $\underline{68.15\!\pm\!6.15}$ & $\underline{68.25\!\pm\!4.64}$ & $\underline{68.25\!\pm\!4.85}$ & $\underline{67.01\!\pm\!5.80}$ \\
Absolute Delta & $\underline{47.71\!\pm\!6.93}$ & $\underline{49.86\!\pm\!12.41}$ & $38.27\!\pm\!2.30$ & $34.69\!\pm\!1.52$ & $62.57\!\pm\!5.74$ & $64.09\!\pm\!6.14$ & $62.57\!\pm\!5.74$ & $61.32\!\pm\!6.36$ & $57.36\!\pm\!9.77$ & $54.24\!\pm\!17.35$ & $60.26\!\pm\!8.78$ & $54.68\!\pm\!13.74$ \\
\midrule
\textbf{Bidirectional Delta} & $\mathbf{55.55\!\pm\!3.96}$ & $\mathbf{54.20\!\pm\!12.81}$ & $\mathbf{48.32\!\pm\!7.60}$ & $\mathbf{47.32\!\pm\!8.21}$ & $\mathbf{83.21\!\pm\!6.25}$ & $\mathbf{84.18\!\pm\!5.50}$ & $\mathbf{83.21\!\pm\!6.25}$ & $\mathbf{83.03\!\pm\!6.41}$ & $\mathbf{84.49\!\pm\!7.36}$ & $\mathbf{85.66\!\pm\!6.83}$ & $\mathbf{83.90\!\pm\!5.85}$ & $\mathbf{83.92\!\pm\!6.93}$ \\
\bottomrule
\end{tabular}%
}
\end{threeparttable}
\end{table*}

%% file: InferenceTime.tex
\begin{table}[ht]
    \centering
    \caption{Model Inference Time Comparison on CPU}
    \label{tab:InferenceTime}
    \begin{tabular}{lrrr}
        \toprule
        Model & Params & FLOPs & Inference Time (ms) \\
        \midrule
        EEGNet          & 3.779K & 15.585M & 4.97   \\
        FCN\_Wang       & 265.091K & 7.217G & 422.39 \\
        GAT\_CNN        & 67.939K & 197.599M & 29.23  \\
        InceptionTime   & 496.547K & 797.133M & 91.31  \\
        ResNet\_Wang    & 539.011K & 863.847M & 56.15  \\
        TSception       & 40.909K & 79.278M & 21.99  \\
        ViT             & 85.240M & 8.607G & 978.45 \\
        XResnet         & 15.968M & 637.346M & 91.72  \\
        LaBraM          & 5.786M & 796.096M & 95.61  \\
        EEGMamba        & 47.522M & 544.133M & 923.95 \\
        \midrule
        DeltaGateNet    & 4.768M & 7.613G & 523.16 \\
        \bottomrule
    \end{tabular}
\end{table}

%% file: Parameter_Tuning.tex
\begin{table*}[t]
\centering
\caption{Kernel Size Tuning Results on SEED-VIG, SADT 2022, and SADT 2952 (5-fold CV)}
\label{tab:combined_kernel_size}
\begin{threeparttable}
\resizebox{\textwidth}{!}{%
\begin{tabular}{@{}lcccccccccccc@{}}
\toprule
\multirow{2}{*}{\textbf{Kernel Size}} & \multicolumn{4}{c}{SEED-VIG} & \multicolumn{4}{c}{SADT 2022} & \multicolumn{4}{c}{SADT 2952} \\
\cmidrule(lr){2-5} \cmidrule(lr){6-9} \cmidrule(lr){10-13}
& Acc & Prec & Rec & F1 & Acc & Prec & Rec & F1 & Acc & Prec & Rec & F1 \\
\midrule
\multicolumn{13}{@{}l}{\textbf{Intrasubject Evaluation}} \\
\midrule
K = 1 & $71.08\!\pm\!3.86$ & $74.39\!\pm\!2.37$ & $68.30\!\pm\!5.48$ & $69.58\!\pm\!4.83$ & $94.92\!\pm\!2.07$ & $94.97\!\pm\!2.02$ & $94.95\!\pm\!2.09$ & $94.90\!\pm\!2.07$ & $95.69\!\pm\!1.76$ & $95.66\!\pm\!1.79$ & $95.50\!\pm\!1.70$ & $95.55\!\pm\!1.75$ \\
K = 3 & $73.04\!\pm\!3.81$ & $77.27\!\pm\!3.57$ & $70.43\!\pm\!5.01$ & $72.12\!\pm\!4.77$ & $96.56\!\pm\!2.08$ & $96.97\!\pm\!2.03$ & $96.95\!\pm\!2.14$ & $96.81\!\pm\!2.59$ & $95.75\!\pm\!1.85$ & $95.65\!\pm\!1.99$ & $95.62\!\pm\!1.84$ & $95.62\!\pm\!1.91$ \\
K = 5 & $74.14\!\pm\!3.74$ & $78.26\!\pm\!3.29$ & $71.87\!\pm\!5.01$ & $73.52\!\pm\!4.68$ & $96.16\!\pm\!2.24$ & $96.18\!\pm\!2.34$ & $96.16\!\pm\!2.23$ & $96.15\!\pm\!2.24$ & $96.70\!\pm\!1.73$ & $96.59\!\pm\!1.87$ & $96.67\!\pm\!1.67$ & $96.61\!\pm\!1.78$ \\
\textbf{K = 7} & $\mathbf{81.89\!\pm\!0.66}$ & $\mathbf{84.37\!\pm\!0.52}$ & $\mathbf{81.32\!\pm\!0.64}$ & $\mathbf{82.55\!\pm\!0.58}$ & $\mathbf{96.81\!\pm\!2.58}$ & $\mathbf{96.98\!\pm\!2.34}$ & $\mathbf{96.80\!\pm\!2.60}$ & $\mathbf{96.81\!\pm\!2.59}$ & $\mathbf{96.84\!\pm\!1.43}$ & $\mathbf{96.73\!\pm\!1.41}$ & $\mathbf{96.81\!\pm\!1.53}$ & $\mathbf{96.77\!\pm\!1.47}$ \\
\midrule
\multicolumn{13}{@{}l}{\textbf{Intersubject Evaluation}} \\
\midrule
K = 1 & $48.73\!\pm\!4.03$ & $49.30\!\pm\!9.92$ & $42.36\!\pm\!5.44$ & $37.72\!\pm\!5.58$ & $65.72\!\pm\!17.05$ & $67.64\!\pm\!20.15$ & $63.34\!\pm\!21.36$ & $62.87\!\pm\!20.17$ & $78.74\!\pm\!6.78$ & $79.76\!\pm\!7.22$ & $78.01\!\pm\!7.99$ & $77.25\!\pm\!8.22$ \\
K = 3 & $48.12\!\pm\!4.27$ & $49.08\!\pm\!11.38$ & $41.00\!\pm\!4.58$ & $37.10\!\pm\!5.30$ & $74.45\!\pm\!8.41$ & $79.30\!\pm\!5.73$ & $74.45\!\pm\!8.41$ & $72.84\!\pm\!9.81$ & $63.88\!\pm\!21.13$ & $64.88\!\pm\!19.83$ & $65.45\!\pm\!20.71$ & $62.13\!\pm\!22.21$ \\
K = 5 & $43.37\!\pm\!4.13$ & $43.54\!\pm\!5.69$ & $40.98\!\pm\!3.15$ & $47.32\!\pm\!8.21$ & $82.25\!\pm\!11.86$ & $88.26\!\pm\!6.88$ & $85.25\!\pm\!11.86$ & $84.10\!\pm\!14.05$ & $77.75\!\pm\!8.82$ & $79.56\!\pm\!6.16$ & $78.51\!\pm\!7.90$ & $76.91\!\pm\!9.02$ \\
\textbf{K = 7} & $\mathbf{55.55\!\pm\!3.96}$ & $\mathbf{54.20\!\pm\!12.81}$ & $\mathbf{48.32\!\pm\!7.60}$ & $\mathbf{47.32\!\pm\!8.21}$ & $\mathbf{83.21\!\pm\!6.25}$ & $\mathbf{84.18\!\pm\!5.50}$ & $\mathbf{83.21\!\pm\!6.25}$ & $\mathbf{83.03\!\pm\!6.41}$ & $\mathbf{84.49\!\pm\!7.36}$ & $\mathbf{85.66\!\pm\!6.83}$ & $\mathbf{83.90\!\pm\!5.85}$ & $\mathbf{83.92\!\pm\!6.93}$ \\
\bottomrule
\end{tabular}%
}
\end{threeparttable}
\end{table*}

\begin{table*}[t]
\centering
\caption{Step Size Tuning Results on SEED-VIG, SADT 2022, and SADT 2952 (5-fold CV)}
\label{tab:combined_step_size}
\begin{threeparttable}
\resizebox{\textwidth}{!}{%
\begin{tabular}{@{}lcccccccccccc@{}}
\toprule
\multirow{2}{*}{\textbf{Step Size}} & \multicolumn{4}{c}{SEED-VIG} & \multicolumn{4}{c}{SADT 2022} & \multicolumn{4}{c}{SADT 2952} \\
\cmidrule(lr){2-5} \cmidrule(lr){6-9} \cmidrule(lr){10-13}
& Acc & Prec & Rec & F1 & Acc & Prec & Rec & F1 & Acc & Prec & Rec & F1 \\
\midrule
\multicolumn{13}{@{}l}{\textbf{Intrasubject Evaluation}} \\
\midrule
\textbf{S = 1} & $\mathbf{81.89\!\pm\!0.66}$ & $\mathbf{84.37\!\pm\!0.52}$ & $\mathbf{81.32\!\pm\!0.64}$ & $\mathbf{82.55\!\pm\!0.58}$ & $\mathbf{96.81\!\pm\!2.58}$ & $\mathbf{96.98\!\pm\!2.34}$ & $\mathbf{96.80\!\pm\!2.60}$ & $\mathbf{96.81\!\pm\!2.59}$ & $\mathbf{96.84\!\pm\!1.43}$ & $\mathbf{96.73\!\pm\!1.41}$ & $\mathbf{96.81\!\pm\!1.53}$ & $\mathbf{96.77\!\pm\!1.47}$ \\
S = 2 & $78.11\!\pm\!2.89$ & $80.82\!\pm\!1.81$ & $77.47\!\pm\!3.57$ & $78.62\!\pm\!3.14$ & $95.60\!\pm\!2.63$ & $95.67\!\pm\!2.53$ & $95.61\!\pm\!2.65$ & $95.60\!\pm\!2.63$ & $95.93\!\pm\!1.76$ & $95.77\!\pm\!1.83$ & $95.92\!\pm\!1.78$ & $95.84\!\pm\!1.81$ \\
S = 3 & $78.10\!\pm\!2.41$ & $80.55\!\pm\!1.69$ & $77.67\!\pm\!2.98$ & $78.69\!\pm\!2.61$ & $95.12\!\pm\!2.79$ & $95.13\!\pm\!2.81$ & $95.16\!\pm\!2.80$ & $95.12\!\pm\!2.80$ & $95.90\!\pm\!0.84$ & $95.89\!\pm\!0.75$ & $95.66\!\pm\!1.05$ & $95.74\!\pm\!0.87$ \\
\midrule
\multicolumn{13}{@{}l}{\textbf{Intersubject Evaluation}} \\
\midrule
\textbf{S = 1} & $\mathbf{55.55\!\pm\!3.96}$ & $\mathbf{54.20\!\pm\!12.81}$ & $\mathbf{48.32\!\pm\!7.60}$ & $\mathbf{47.32\!\pm\!8.21}$ & $\mathbf{83.21\!\pm\!6.25}$ & $\mathbf{84.18\!\pm\!5.50}$ & $\mathbf{83.21\!\pm\!6.25}$ & $\mathbf{83.03\!\pm\!6.41}$ & $\mathbf{84.49\!\pm\!7.36}$ & $\mathbf{85.66\!\pm\!6.83}$ & $\mathbf{83.90\!\pm\!5.85}$ & $\mathbf{83.92\!\pm\!6.93}$ \\
S = 2 & $49.16\!\pm\!7.14$ & $53.21\!\pm\!10.90$ & $44.43\!\pm\!8.11$ & $41.30\!\pm\!6.49$ & $75.42\!\pm\!8.95$ & $78.00\!\pm\!7.02$ & $75.42\!\pm\!8.95$ & $74.42\!\pm\!10.05$ & $81.66\!\pm\!10.16$ & $82.23\!\pm\!10.38$ & $81.62\!\pm\!10.00$ & $81.22\!\pm\!10.11$ \\
S = 3 & $47.51\!\pm\!8.76$ & $50.25\!\pm\!11.32$ & $43.16\!\pm\!7.15$ & $39.33\!\pm\!7.19$ & $81.28\!\pm\!12.27$ & $82.60\!\pm\!12.04$ & $81.28\!\pm\!12.27$ & $81.03\!\pm\!12.38$ & $81.70\!\pm\!12.25$ & $83.53\!\pm\!9.42$ & $81.90\!\pm\!10.49$ & $79.69\!\pm\!12.39$ \\
\bottomrule
\end{tabular}%
}
\end{threeparttable}
\end{table*}